\documentclass[preprint,12pt,authoryear]{elsarticle}

\usepackage[left=3cm,top=3cm,right=3cm,bottom=3cm]{geometry}
\usepackage{amssymb}
\usepackage{amsmath}
\usepackage{amsfonts}
\usepackage{mathtools}
\usepackage{mathrsfs}
\usepackage{nicefrac}
\usepackage[section]{placeins}
\usepackage{booktabs}
\usepackage{tabularx,multirow}
\usepackage{longtable}
\usepackage{graphicx}
\usepackage{floatrow}
\usepackage{threeparttable}
\usepackage{subfig}
\usepackage{caption}
\usepackage{url}
\usepackage[usenames,dvipsnames]{xcolor}
\usepackage[colorlinks=true, linkcolor=blue, urlcolor=blue, citecolor=blue, anchorcolor=blue]{hyperref}
\usepackage{xr} 
\usepackage{natbib}
\usepackage{appendix}
\usepackage{makecell}

\hypersetup{colorlinks=true,linkcolor={blue},citecolor={blue},filecolor={blue},urlcolor={blue}} 

\makeatletter
\def\ps@pprintTitle{%
	\let\@oddhead\@empty
	\let\@evenhead\@empty
	\def\@oddfoot{\reset@font\hfil\thepage\hfil}
	\let\@evenfoot\@oddfoot
}
\makeatother

\begin{document}

\begin{frontmatter}

\title{Modelling and short-term forecasting of seasonal mortality}

\author{Ainhoa-Elena Leger\textsuperscript{a*}, Silvia Rizzi\textsuperscript{a}, Ugofilippo Basellini\textsuperscript{b}}
\address{\textsuperscript{a}University of Southern Denmark (Cpop), Campusvej 55, 5230 Odense M, Denmark}
\address{\textsuperscript{*}Corresponding author, aleg@sdu.dk}
\address{\textsuperscript{b}Max Planck Institute for Demographic Research, Konrad-Zuse-Str. 1, 18057 Rostock, Germany}

\begin{abstract}
Excess mortality, i.e. the difference between expected and observed mortality, is used to quantify the death toll of mortality shocks, such as infectious disease-related epidemics and pandemics. However, predictions of expected mortality are sensitive to model assumptions. Among three specifications of a Serfling-Poisson regression for seasonal mortality, we analyse which one yields the most accurate predictions. We compare the Serfling-Poisson models with: 1) parametric effect for the trend and seasonality (SP), 2) non-parametric effect for the trend and seasonality (SP-STSS), also known as modulation model, and 3) non-parametric effect for the trend and parametric effect for the seasonality (SP-STFS). Forecasting is achieved with P-splines smoothing. The SP-STFS model resulted in more accurate historical forecasts of monthly rates from national statistical offices in 25 European countries. An application to the COVID-19 pandemic years illustrates how excess mortality can be used to evaluate the vulnerability of populations and aid public health planning.
\end{abstract}

\begin{keyword}
demographic forecasting \sep evaluating forecasts \sep baseline estimation \sep smoothing techniques \sep seasonality
\end{keyword}

\end{frontmatter}

\section{Introduction}
\label{section1}
Measuring the mortality burden related to natural and health shocks provides fundamental information to aid public health responses, by guiding policy-making decisions. It provides insights into the vulnerability of populations, the geographical gradient, and the effect of the policies adopted in response to the shocks. Excess mortality is a useful indicator to assess the impact of influenza outbreaks \citep{mazick2012excess, molbak2015excess, nielsen2018influenza}, heat waves \citep{fouillet2006excess, toulemon2008mortality} and pandemics caused by infectious diseases, such as the 1918–1920 H1N1 influenza pandemic \citep{ansart2009mortality} and the 2019 coronavirus disease (COVID-19) pandemic \citep{kontis2020magnitude, islam2021excess}.

Excess mortality is computed as the difference between the expected and reported mortality in the same period. The expected mortality (or baseline mortality) is predicted in a counterfactual scenario where no mortality shock has happened. The mortality difference, \textit{ceteris paribus}, can be considered the overall effect of the mortality shock. The predictions of the expected mortality are usually obtained in the short term, i.e., within an epidemic year, to study the effect of the influenza season, or over a few years, in the case of several waves of new infectious disease-related pandemics.

The methodological choices about how to forecast mortality are crucial, given that different models lead to varying estimates of excess mortality. Various models were proposed to estimate the expected mortality, which show considerable seasonal variation, in Europe mostly striking harder in winter than summer. The first attempt to model a “standard curve of expected seasonal mortality”, accounting for long-time trends and seasonal variation, was conducted by \cite{serfling1963methods}. This method is based on a linear regression that models the mortality during an epidemiological year as a linear function of time (long-time trend) and one or more sinusoidal terms (seasonal variation). More recently employed versions of the Serfling model are a Poisson regression \citep{thompson2009estimates} considering the count nature of mortality data, and a Quasi-Poisson regression accounting for over-dispersion \citep{euromomo2017european}.

Although widely used, a limitation of the Serfling-Poisson model is the assumption of the linearity of the trend on the logarithmic scale, which can lead to the under or over-estimation of excess mortality, if the trend is not linear. \cite{eilers2008modulation} introduced the modulation models that relax the linearity assumption, by considering smooth long-term trends and smooth seasonal effects over time. The models use regression splines, specifically B-splines with penalties, known as P-splines \citep{eilers1996flexible}. The regression with P-splines leads to a generalised linear model, which is fitted by penalised likelihood. Our aim in this study is to extend the modulation models for seasonal mortality forecasting purposes. We do so by following the approach of \cite{currie2004smoothing}. This approach is usually used within the scope of forecasting death rates over extended periods, such as 50 years, or from a cohort perspective \citep{kirkby2010smooth, camarda2019smooth}. This method integrates P-spline regression with a missing value approach. The same approach can be used to forecast seasonal mortality in the short term. Future values are considered missing values and estimated simultaneously with the fitting of the mortality model. By modifying the weights in the penalisation, forecasting is a natural consequence of the smoothing process.

In this study, we consider the prediction of the baseline mortality when no mortality shock happens. Our study builds on the works of \cite{eilers2008modulation} and \cite{currie2004smoothing}. The Serfling-Poisson model is used as the basis to forecast baseline mortality in the short term via P-splines, with a missing value approach. The P-spline smoothing is adapted to predict the expected mortality during one or more epidemiological years. This yields the baseline mortality used in estimating excess mortality during influenza outbreaks or mortality shocks. We compare 1) the classical Serfling-Poisson model and two model versions with: 2) smooth long-term trends and varying seasonal components (also called a modulation model), and 3) smooth long-term trends and fixed seasonality. For our application of the models, we retrieved the mortality and population data on 25 European countries from Eurostat.

In Section \ref{section2}, we describe our proposed forecasting strategy. We start by reviewing the Serfling-Poisson model, then show the two proposed variants of the model, illustrating the estimation and forecasting strategy. In Section \ref{section3}, we describe our data sets. In Sections \ref{section4} and \ref{section5}, we describe our choices for the models' parameters, respectively, for mortality modelling and mortality forecasting. We show an application to the estimation of excess mortality during the COVID-19 pandemic. The paper concludes with a critical discussion of our methodology and the findings for the excess mortality estimation.
\section{Methods}
\label{section2}
This section first presents the model traditionally used in the literature to predict seasonal mortality (subsection \ref{section2.1}) and the proposed models with a detailed description of the regression matrix and penalty matrix (subsections \ref{section2.2} and \ref{section2.3}). The method to estimate and forecast with P-splines is then adapted to seasonal data (subsection \ref{section2.4} and \ref{section2.5}). Finally, we explain the measures that will be used to evaluate the model's goodness of fit and the forecasts' accuracy (subsection \ref{section2.6}).
\subsection{Serfling-Poisson (SP) model}
\label{section2.1}
Let $Y_{t}$ be a non-negative random variable denoting the death counts in a population at the months $t$, with $t=1,...,T$. The realisations of $Y_{t}$ are the observed number of deaths $y_{t}$. We assume that the random variable $Y_{t}$ follows a Poisson distribution with expected values $\mu_{t}$.
\begin{equation*}
	Y_{t} \sim Poi(\mu_{t}), \quad \mu_{t} = E[Y_{t}]
\end{equation*}
The log link function relates the mean $\mu_{t}$ to the linear predictor $log(\mu_{t}) = \eta_{t}$. \\
The first model we consider is the Serfling-Poisson model \citep{serfling1963methods}, which includes a linear trend and models the seasonality using sine and cosine functions.
\begin{equation*}
	log(\mu_{t}) = \beta_{0} + \beta_{1}t + \beta_{3}cos(wt) + \beta_{4}sin(wt)
\end{equation*}	
where $t=1,\dots,T$, $w=2\pi/p$, and $p$ is the period. Analyses in this paper are performed on monthly death counts and rates, so $p=12$. Estimation of the regression coefficients $\hat{\beta}$ can be performed with the Iterated Weighted Least Squares (IWLS) for Generalized Linear Models (GLMs) \citep{mccullagh2019generalized}.

The model allows for exposures $e_{t}$, when the objective is to model death rates. The Serfling-Poisson (SP) with exposures is
\begin{equation*}
	log(\mu_{t}) = log(e_{t}) + \beta_{0} + \beta_{1}t + \beta_{3}cos(wt) + \beta_{4}sin(wt).
\end{equation*}	
\subsection{Smooth trend and smooth seasonality (SP-STSS) model}
\label{section2.2}
Secondly, we propose to use the modulation models developed by \cite{eilers2008modulation} and extend them to forecast the baseline mortality, by adapting the P-splines approach of \cite{currie2004smoothing} to seasonal data. The modulation models introduce a smooth trend function and time-varying coefficients. This gives a very general model for demographic seasonal time series. The structure is the following 
\begin{equation}
	\label{eq:mm}
	log(\mu_{t}) = \upsilon_{t} + f_{t}cos(wt) + g_{t}sin(wt),
\end{equation}
where $\upsilon_{t}$ accounts for the smooth trend, $f_{t}$ and $g_{t}$ are smooth functions that describe the local amplitudes of the cosine and sine waves, and $w=2\pi/p$, with $p$ the period (e.g., $p=12$ for monthly data). The smooth trend function and the modulation series $f_{t}$ and $g_{t}$ are constructed by approximating B-spline basis. Specifically, $\upsilon_{t}=\sum_{j}\alpha_{j}B_{j}(t)$, $f_{t}=\sum_{j}\beta_{j}B_{j}(t)$ and $g_{t}=\sum_{j}\gamma_{j}B_{j}(t)$, with $\boldsymbol{B}=[b_{tj}]=[B_{j}(t)]$ a basis of B-splines, $t=1,\dots,T$ the time index, and $j=1,\dots,J$ the B-splines index.

By introducing the matrices $\boldsymbol{C} = diag\{cos(wt)\}$ and $\boldsymbol{S} = diag\{sin(wt)\}$, the model in Equation \ref{eq:mm} can be written in the matrix-vector notation
\begin{equation*}
	log(\boldsymbol{\mu}) = \boldsymbol{B\alpha} + \boldsymbol{CB \beta} + \boldsymbol{SB \gamma} = \boldsymbol{\eta}
\end{equation*}
where $\boldsymbol{\upsilon} = \boldsymbol{B\alpha}$, $\boldsymbol{f} = \boldsymbol{B \beta}$, $\boldsymbol{g} = \boldsymbol{B \gamma}$. The linear predictor can then be re-arranged 
\begin{equation}
	\label{eq:mm_glm}
	\eta = [\boldsymbol{B}|\boldsymbol{CB}|\boldsymbol{SB}] [\boldsymbol{\alpha'}|\boldsymbol{\beta'}|\boldsymbol{\gamma'}] = \boldsymbol{\breve{B}\theta}. \footnote{From here onward, $\boldsymbol{B}$ indicates the matrix of B-splines, while $\boldsymbol{\breve{B}}$ indicates the regression matrix.}
\end{equation}
When modelling death rates, the exposures $e_{t}$ are included and the SP-STSS model becomes
\begin{equation*}
	log(\mu_{t}) = log(e_{t}) + \upsilon_{t} + f_{t}cos(wt) + g_{t}sin(wt),
\end{equation*}
\subsection{Smooth trend and fixed seasonality (SP-STFS) model}
\label{section2.3}
In addition to employing the Serfling-Poisson and modulation models, we propose an alternative approach to account for a smooth trend and fixed seasonality (STFS). Specifically, we employ the same structure
\begin{equation*}
	log(\mu_{t}) = \upsilon_{t} + \beta_{1}cos(wt) + \beta_{2}sin(wt)
\end{equation*}
where $\upsilon_{t}$ accounts for the smooth trend, and the cosine and sine have constant coefficients $\beta_{1}$ and $\beta_{2}$ over time, and $w=2\pi/p$. The smooth trend function is constructed by B-spline approximation. Specifically, $\upsilon_{t}=\sum_{j}\alpha_{j}B_{j}(t)$, with $\boldsymbol{B}=[b_{tj}]=[B_{j}(t)]$ a basis of B-splines, $t=1,\dots,T$ the time index, and $j=1,\dots,J$ the B-splines' index.

The linear predictor models the trend component with the B-splines matrix $\boldsymbol{B}$ and time-varying coefficients $\boldsymbol{\alpha}$, and the seasonal component with the vectors $\boldsymbol{c}=cos(wt)$ and $\boldsymbol{s}=sin(wt)$ and the coefficients $\beta_{1}$ and $\beta_{2}$
\begin{equation}
	\label{eq:sm_glm}
	\boldsymbol{\eta} = [\boldsymbol{B}|\boldsymbol{c'}|\boldsymbol{s'}] [\boldsymbol{\alpha'}|\beta_{1}|\beta_{2}] = \boldsymbol{\breve{B}\theta}.
\end{equation}
When modelling death rates, the exposures $e_{t}$ are included and the SP-STFS model becomes
\begin{equation*}
	log(\mu_{t}) = log(e_{t}) + \upsilon_{t} + \beta_{1}cos(wt) + \beta_{2}sin(wt).
\end{equation*}
\subsection{Estimation}
\label{section2.4}
For both the SP-STSS model and the SP-STFS model, the estimation of the model's parameters is achieved using penalised B-splines or P-splines, to force them to vary more smoothly \citep{eilers1996flexible}. The B-spline bases model the series $v$, $f$ and $g$ and an additional penalty on the B-spline coefficients optimises their amount of smoothing.

We minimise the penalised Poisson deviance defined as
\begin{equation*}
	d^{*}(y;\mu) = 2 \sum_{t=1}^{T} log(y_{t}/\mu_{t})+ \lambda_{1} \left\lVert \boldsymbol{D\alpha} \right\rVert ^2 + \lambda_{2} \left\lVert \boldsymbol{D\beta} \right\rVert ^2 + \lambda_{2} \left\lVert \boldsymbol{D\gamma} \right\rVert ^2
\end{equation*}
where the matrix $\boldsymbol{D}=\Delta^{d}$ constructs $d$th order differences of $\boldsymbol{\alpha}$, $\boldsymbol{\beta}$ and $\boldsymbol{\gamma}$. For instance, $\Delta^{1}$ is a matrix $(J-1)\times J$ of first differences and $\boldsymbol{\Delta^{1}\alpha}$ is the vector with elements $\boldsymbol{\alpha_{j+1}}-\boldsymbol{\alpha_{j}}$, for $j=1,\dots,J-1$. By repeating this computation on $\boldsymbol{\Delta^{1}\alpha}$, we arrive at higher differences like $\boldsymbol{\Delta^{2}\alpha}$, where $\boldsymbol{\Delta^{2}}$ the $(J-2)\times J$ matrix of second-order differences of a J-vector.

The linear re-expressions \ref{eq:mm_glm} and \ref{eq:sm_glm} allow all of the parameters associated with each component to be estimated simultaneously as GLMs. Estimation of the coefficients is performed via the penalized version of the Iterated Weighted Least Squares (IWLS)
\begin{equation}
	\label{eq:iwls}
	(\boldsymbol{\breve{B}'} \boldsymbol{\widetilde{M}^{(t)}} \boldsymbol{\breve{B}} + \boldsymbol{P}) \boldsymbol{\theta^{(t+1)}} = \boldsymbol{\breve{B}'} \boldsymbol{\widetilde{M}^{(t)}} \boldsymbol{\breve{B}} \boldsymbol{\theta^{(t)}} + \boldsymbol{\breve{B}'} (\boldsymbol{y}-\boldsymbol{\tilde{\mu}}),
\end{equation}
where $\boldsymbol{\breve{B}}$ is the regression matrix, $\boldsymbol{\widetilde{M}} = diag(\boldsymbol{\mu})$ is the matrix of weights, and $\boldsymbol{P} = \Lambda \boldsymbol{D'} \boldsymbol{D}$ is the penalty term. The positive penalty hyper-parameter $\Lambda = diag(\lambda_{1}, \lambda_{2}, \lambda_{2})$ balances smoothness against fit to the data and allows for different penalties for the trend ($\lambda_{1}$) and modulation functions ($\lambda_{2}$). The penalty matrix can also be constructed to consider different orders of the differences for the trend and modulation functions as $\boldsymbol{P}=blockdiag(\lambda_{1}\boldsymbol{D_{1}'}\boldsymbol{D_{1}},\lambda_{2}\boldsymbol{D_{2}'}\boldsymbol{D_{2}},\lambda_{2}\boldsymbol{D_{2}'}\boldsymbol{D_{2}})$. For instance, \cite{carballo2021} suggest an order of differences of 2 and 1 for $\lambda_{1}$ and $\lambda_{2}$, respectively.

The modulation model presents some similarities with some other models in the literature. The smooth trend component $\boldsymbol{\upsilon}$ can be seen as a generalised additive model (GAM, \citealt{hastie1990}) and the seasonal components $\boldsymbol{f}$ and $\boldsymbol{g}$ as a varying-coefficient model (VCM, \citealt{hastie1993}). The advantage of P-splines over GAM and VCM is that they avoid both the backfitting algorithm and complex knot selection schemes.
\subsection{Forecasting with P-splines}
\label{section2.5}
Following \cite{currie2004smoothing}, forecasting future values can be treated as a missing value problem, and the fitted and forecast values can be estimated simultaneously. Let us consider the observed death counts $y_{1}$ for $n_{1}$ months. We want to forecast $n_{2}$ months into the future. Let us define a new time index $t_{+}$ with $t_{+}=1,...,n_{1},...,n_{1}+n_{2}$. To obtain the fit and the forecast simultaneously, we extend for $n_{1}+n_{2}$ months the regression matrix $\boldsymbol{\breve{B}}$ and the penalty matrix $\boldsymbol{P}$. The IWLS algorithm of Equation \ref{eq:iwls} is adapted as follows: 
\begin{equation*}
	(\boldsymbol{\breve{B}_{+}'} \boldsymbol{V} \boldsymbol{\widetilde{M}^{(t)}} \boldsymbol{\breve{B}_{+}} + \boldsymbol{P_{+}}) \boldsymbol{\theta^{(t+1)}} = \boldsymbol{\breve{B}_{+}'} \boldsymbol{V} \boldsymbol{\widetilde{M}^{(t)}} \boldsymbol{\breve{B}_{+}} \boldsymbol{\theta^{(t)}} + \boldsymbol{\breve{B}_{+}'} \boldsymbol{V} (\boldsymbol{y}-\boldsymbol{\tilde{\mu}}),
\end{equation*}
where the matrix $\boldsymbol{\breve{B}_{+}}=[\boldsymbol{\breve{B}_{+}}|\boldsymbol{C_{+}B_{+}}|\boldsymbol{S_{+}B_{+}}]$ is the extended regression matrix, and the matrix $\boldsymbol{P_{+}}=blockdiag(\lambda_{1}\boldsymbol{D_{1+}'}\boldsymbol{D_{1+}},\lambda_{2}\boldsymbol{D_{2+}'}\boldsymbol{D_{2+}},\lambda_{2}\boldsymbol{D_{2+}'}\boldsymbol{D_{2+}})$ is the extended penalty matrix. It follows that $\boldsymbol{\breve{B}_{+}}$ is the extended matrix of B-splines for the trend, $\boldsymbol{C_{+}}=diag(cos(wt_{+}))$ and $\boldsymbol{S_{+}}=diag(sin(wt_{+}))$ are the extended matrices for the modulation components, and $\boldsymbol{D_{1+}}$ and $\boldsymbol{D_{2+}}$ are the extended matrices of differences. Furthermore, a matrix $\boldsymbol{V}=blockdiag(\boldsymbol{I};\boldsymbol{0})$ weights 1 the observations ($\boldsymbol{I}$ has size $n_{1}$) and 0 the forecasts ($\boldsymbol{0}$ has size $n_{2}$).

The confidence intervals for the fitted values and the prediction intervals (PIs) for the forecasts are computed simultaneously using the approximate variance of $\boldsymbol{\breve{B}_{+}\hat{\theta}}$ given by
\begin{equation*}
	Var(\boldsymbol{\breve{B}_{+}}\hat{\theta}) \approx \boldsymbol{\breve{B}_{+}}(\boldsymbol{\breve{B}_{+}'} \boldsymbol{V_{+}} \boldsymbol{\widetilde{M}} \boldsymbol{\breve{B}_{+}} + \boldsymbol{P_{+}})^{-1}\boldsymbol{\breve{B}_{+}'}.
\end{equation*}
\subsection{Evaluation measures}
\label{section2.6}
To evaluate the model performance balancing the goodness of fit and the model complexity, we will use the Bayesian Information Criterion (BIC) defined as \citep{schwarz1978estimating}
\begin{equation*}
	\text{BIC} = 2 \text{Dev} + \text{log n Tr}
\end{equation*}	
where Dev is the model's deviance, $n$ is the number of observations, and $Tr=tr(\boldsymbol{H})$ is the effective dimension of the fitted model. The hat matrix is defined as $\boldsymbol{H}=\boldsymbol{\hat{M}}^{1/2} \boldsymbol{\breve{B}} (\boldsymbol{\breve{B}}' \boldsymbol{\hat{M}} \boldsymbol{\breve{B}} + \boldsymbol{P})^{-1} \boldsymbol{\breve{B}'} \boldsymbol{\hat{M}}^{1/2}$ and the trace is $tr(\boldsymbol{H}) = tr[\boldsymbol{\breve{B}}' \boldsymbol{\hat{M}} \boldsymbol{\breve{B}} (\boldsymbol{\breve{B}'} \boldsymbol{\hat{M}} \boldsymbol{\breve{B}} + \boldsymbol{P}) ^{-1}]$.

To evaluate the forecasting accuracy, we will use the Root Mean Squared Error (RMSE) and the Mean Absolute Percentage Error (MAPE). We consider $y_{1},\dots,y_{N}$ to fit the model (training set) and $y_{N+1},\dots,y_{T}$ to check the forecasting accuracy (test set). The forecasting error is defined as the difference between the observed values and the forecasts $e_{t}=y_{t}-\hat{y}_{t}$, for $t=N+1,\dots,T$ 
\begin{equation*}
	RMSE = \sqrt{mean(e^{2}_{t})}=\frac{1}{T-N} \sum_{t=N+1}^{T} e_{t}^2.
\end{equation*}
The percentage error is defined as $p_{t}=100 e_{t}/{y}_{t}$, for $t=1,\dots,T$
\begin{equation*}
	MAPE = mean(\lvert p_{t} \rvert)=\frac{1}{T-N} \sum_{t=N+1}^{T} \lvert p_{t} \rvert.
\end{equation*}
\section{Data}
\label{section3}
We retrieved the series of monthly death counts and population counts from Eurostat\footnote{\url{https://ec.europa.eu/eurostat}} for 25 countries (Austria, Bulgaria, Croatia, Czechia, Denmark, Estonia, Finland, France, Germany, Greece, Hungary, Iceland, Ireland, Italy, Lithuania, Luxembourg, the Netherlands, Norway, Poland, Portugal, Romania, Slovenia, Spain, Sweden, and Switzerland). We selected countries where we could access a complete series of monthly data that go back to at least 1995 and extend through June 2022.

We use the yearly population counts to obtain the exposures as the mid-point between populations on the 1st of January of each year and divided by 12. The deaths and exposure data are arranged, respectively, in vectors $\boldsymbol{y}$ and $\boldsymbol{e}$, indexed by month. The crude death rates (CDRs) are computed as $\boldsymbol{y/e}$ and measure the intensity of deaths in a population and are used to project mortality in time and compute a baseline mortality \citep{aburto2021estimating, basellini2021linking, stokes2021covid}.

\begin{figure}[!t]
	\includegraphics[width=15cm,keepaspectratio]{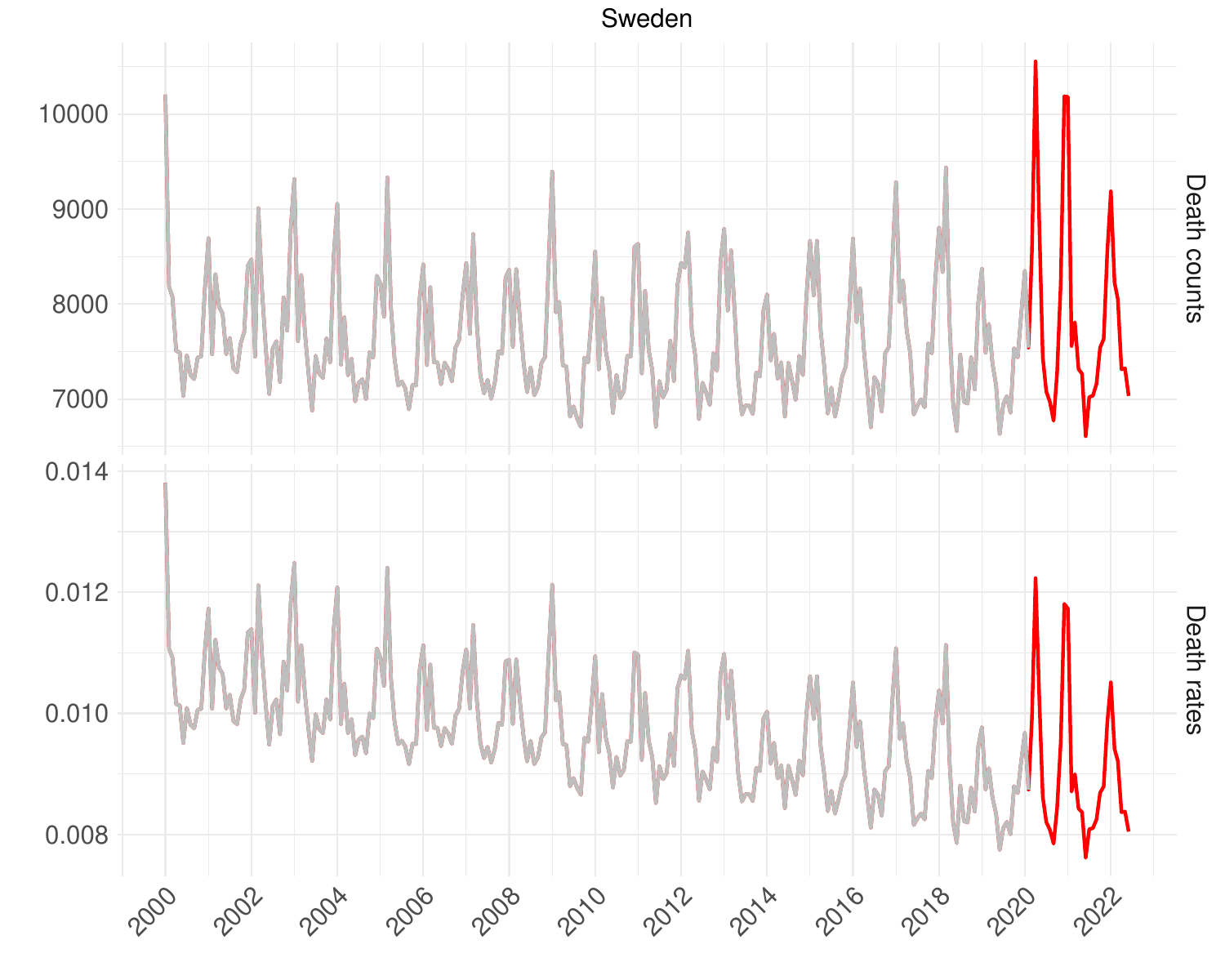}
	\caption{\textbf{Death counts and CDRs from 2010 to 2022 in Sweden}. The period before the COVID-19 pandemic is grey (2009 to February 2020), and the period after the COVID-19 pandemic is red (March 2020 to June 2022).}
	\label{fig:fig_data}
\end{figure}
Figure \ref{fig:fig_data} shows, for illustrative purposes, the monthly total death counts and CDRs for Sweden from 2000 to 2022. The series show a strong temporal trend and cyclical behavior, also known as seasonal variation. The CDRs decrease over time, influenced by improved living conditions and mortality. The seasonal variation strikes harder in winter than in summer, as it is driven by seasonal influenza in the older population. The winter peak usually happens every year between December and March. In February 2020, the COVID-19 pandemic struck in Europe, disrupting mortality patterns and increasing mortality outside the usual window of December to March. In the two years after the first wave of COVID-19, mortality seems to follow the regular pattern of seasonal mortality.

Additionally, the series were retrieved, disaggregated by sex and age at death, from the national statistical offices\footnote{Statistics Denmark, \url{https://www.dst.dk/en}, Istituto Nazionale di Statistica \url{https://www.istat.it/}, Instituto Nacional de Estadística, \url{https://www.ine.es/en/}, Statistics Sweden, \url{https://www.scb.se/en/}} of the countries that made them available: Denmark (2007 to 2022), Italy (2003 to 2022), Spain (2009 to 2022), and Sweden (2000 to 2022). We grouped the death counts by four age groups (0-64, 65-74, 75-84, 85+) that are comparable across the four countries. When comparing mortality across populations, considering the age-specific death rates (ASDRs) instead of the CDRs reduces the influence of the age composition of the population \citep{nemeth2021open, nepomuceno2022sensitivity}. Age structures can change over time, in relation to, for instance, population ageing \citep{preston2001demography, djeundje2022slowdown, missov2023improvements}.
\section{The underlying models}
\label{section4}
The previous section set out the theory for modelling with P-splines. In this section, we explain our choice of the various parameters in the modelling (subsection \ref{section4.1}). We then compare the performance of the three models (SP, SP-STSS, SP-STFS) in all European countries from 1995 to 2019, i.e., in the absence of the mortality shock caused by the COVID-19 pandemic (subsection \ref{section4.2}).
\subsection{Choice of the models' parameters}
\label{section4.1}
The parameters to be chosen when fitting the SP-STSS and SP-STFS models are: 1) the degree of the splines, 2) the number of regions (or equivalently the number of knots) to divide the domain, and 3) the penalty hyper-parameters.

The splines' degree controls the order of the polynomial included in the B-spline basis. We used a basis of cubic B-splines (degree 3).

The chosen number of regions to divide the domain is two per year. Therefore, the number of B-splines depends on the number of years used to predict the baseline. The number of B-splines is the number of divisions plus the degree of the splines. The relation can also be expressed as $ndx=nknots-1+bdeg$, with $ndx$ number of basis functions, $nknots$ number of knots, and $bdeg$ degree of the spline. For instance, let us fit the model on 10 years (from 2010 to 2019): the domain is divided into 20 regions by 21 knots (19 interior knots and 2 knots at the boundaries of the domain) and uses 23 cubic B-splines.

\begin{figure}[h!t]
	\begin{center}
		\includegraphics[width=14cm,keepaspectratio]{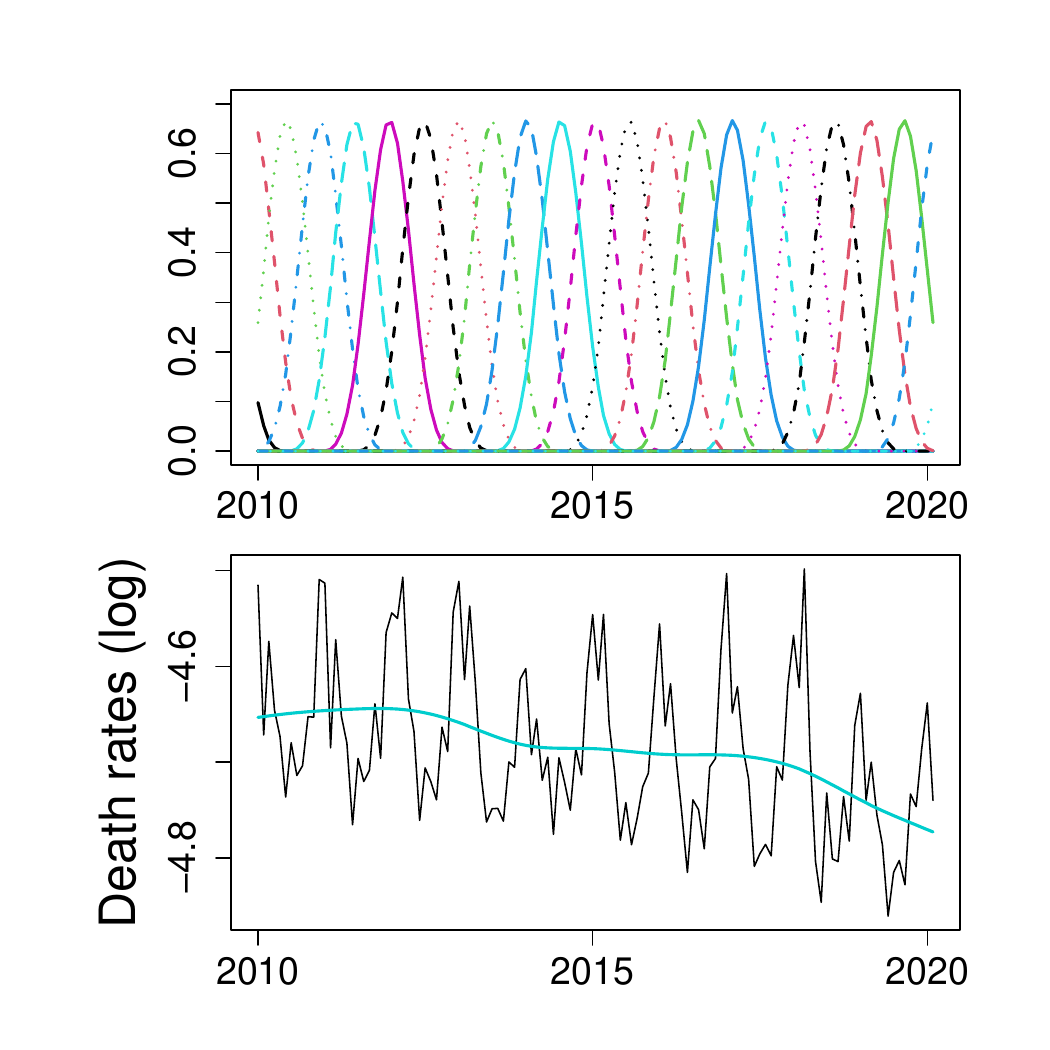}
		\caption{\textbf{Illustration of the fit (logarithmic scale) of the trend function with B-splines for Sweden}. The top panel shows a basis of 23 cubic B-splines, and the bottom panel shows the sum of the scaled B-spline basis functions.}
		\label{fig:fig_basis}
	\end{center}
\end{figure}
Figure \ref{fig:fig_basis} illustrates the non-parametric fit of the trend function $v_{t}$ with B-splines for Sweden. The results are presented on a logarithmic scale, because the Poisson regression models the logarithm of the expected value. The top panel shows a basis of 23 cubic B-splines. In the bottom panel, the B-splines locally defined are scaled with the estimated coefficients (e.g., negative coefficients lead to parabolas below zero; the higher the coefficient, the higher the scaled parabola) to fit the monthly CDRs (black line). The scaled B-spline functions are then summed up to form the smooth trend function $\boldsymbol{\upsilon} = \boldsymbol{B\alpha}$ that fits the data (light blue line). The B-splines are then smoothly joined together by construction.

The P-splines introduce an additional penalty term that prevents overfitting. In the SP-STSS model, the penalty uses second order differences for the trend and first-order differences for the seasonal component \citep{carballo2021}. In the SP-STFS model, we use second-order differences for the trend. The tuning of the penalty hyper-parameters $\lambda_{1}$ for the trend, and $\lambda_{2}$ for the seasonal component was achieved through MAPE minimisation on a grid search. We will explain these choices in detail in subsection \ref{section5.1}, because they influence the forecasts.
\subsection{Performance of the fitted models}
\label{section4.2}
To compare their goodness of fit, the three models (SP, SP-STSS and SP-STFS) were fitted on a rolling window, and the model fit was measured with the BIC. We used a 5-year rolling window because it is the standard choice in the literature, and a 10-year rolling window as a comparison. The three models were fitted on the common period available for the death series, that is from 2010 to 2019. The series' lengths permit the computation of the BIC for a sufficient number of years for both the 5-year rolling window (20 BICs) and the 10-year rolling window (15 BICs). Table \ref{tab:bic} shows the mean BIC values obtained by fitting the models on the CDRs. The mean BIC favours the model SP-STSS, which allows for more flexibility of the trend and seasonality, in almost all the countries (20 countries with the 5-year rolling window and 21 countries with the 10-year rolling window). Either using 5 years or 10 years for the rolling window, the SP-STSS model fits the series better than either the SP or the SP-STFS model. The results of fitting the three models to the death counts are shown in \ref{appendixA} (Table \ref{tab:bic2}) and are similar to the results for the CDRs.

\begin{table}[t]
\begingroup\fontsize{8}{10}\selectfont
\begin{longtable}[c]{>{}l|cc>{}c|ccc}
	\caption{\label{tab:bic}\textbf{Mean BIC on CDRs in 25 European countries for multiple fitting periods based on a rolling-window scheme}. Three models (SP, SP-STSS, and SP-STFS) and two lengths of the fitting period are compared (5 and 10 years).}\\
	\toprule
	\multicolumn{1}{c}{ } & \multicolumn{3}{c}{5 years series} & \multicolumn{3}{c}{10 years series} \\
	\cmidrule(l{3pt}r{3pt}){2-4} \cmidrule(l{3pt}r{3pt}){5-7}
	Country & SP & SP-STSS & SP-STFS & SP & SP-STSS & SP-STFS\\
	\midrule
	Austria & 1,032 & 987 & 1,013 & 2,038 & 1,971 & 2,000\\
	Bulgaria & 1,895 & 1,836 & 1,847 & 3,809 & 3,671 & 3,722\\
	Croatia & 836 & 800 & 822 & 1,641 & 1,605 & 1,633\\
	Czechia & 1,322 & 1,272 & 1,295 & 2,511 & 2,352 & 2,379\\
	Denmark & 648 & 642 & 653 & 1,346 & 1,305 & 1,327\\
	\addlinespace
	Estonia & 243 & 252 & 253 & 487 & 488 & 490\\
	Finland & 597 & 597 & 608 & 1,214 & 1,203 & 1,218\\
	France & 8,099 & 7,403 & 7,867 & 17,036 & 15,934 & 16,448\\
	Germany & 10,556 & 9,582 & 10,099 & 21,869 & 20,058 & 20,837\\
	Greece & 2,646 & 2,510 & 2,576 & 5,066 & 4,871 & 4,969\\
	\addlinespace
	Hungary & 2,163 & 2,056 & 2,124 & 4,143 & 4,046 & 4,090\\
	Iceland & 96 & 108 & 108 & 198 & 214 & 214\\
	Ireland & 686 & 679 & 694 & 1,495 & 1,384 & 1,432\\
	Italy & 11,248 & 10,008 & 10,556 & 22,304 & 20,579 & 21,429\\
	Lithuania & 599 & 586 & 592 & 1,463 & 1,246 & 1,254\\
	\addlinespace
	Luxembourg & 107 & 119 & 119 & 216 & 231 & 231\\
	Netherlands & 1,853 & 1,773 & 1,813 & 3,753 & 3,596 & 3,616\\
	Norway & 618 & 592 & 626 & 1,248 & 1,173 & 1,240\\
	Poland & 4,866 & 4,501 & 4,704 & 9,776 & 8,974 & 9,457\\
	Portugal & 3,529 & 3,392 & 3,489 & 6,806 & 6,691 & 6,756\\
	\addlinespace
	Romania & 4,199 & 3,759 & 3,861 & 9,192 & 8,172 & 8,676\\
	Slovenia & 314 & 315 & 323 & 637 & 640 & 649\\
	Spain & 9,435 & 8,614 & 9,026 & 19,226 & 18,056 & 18,568\\
	Sweden & 1,094 & 1,067 & 1,096 & 2,218 & 2,201 & 2,225\\
	Switzerland & 813 & 783 & 814 & 1,600 & 1,590 & 1,606\\
	\addlinespace
	No. countries * & 4 & 20 & 1 & 4 & 21 & 0\\
	\bottomrule
	\multicolumn{7}{l}{\rule{0pt}{1em}\textsuperscript{*} Number of countries in which the model performs the best}\\
\end{longtable}
\endgroup{}
\end{table}

\begin{figure}[h!p]
	\begin{center}
		\includegraphics{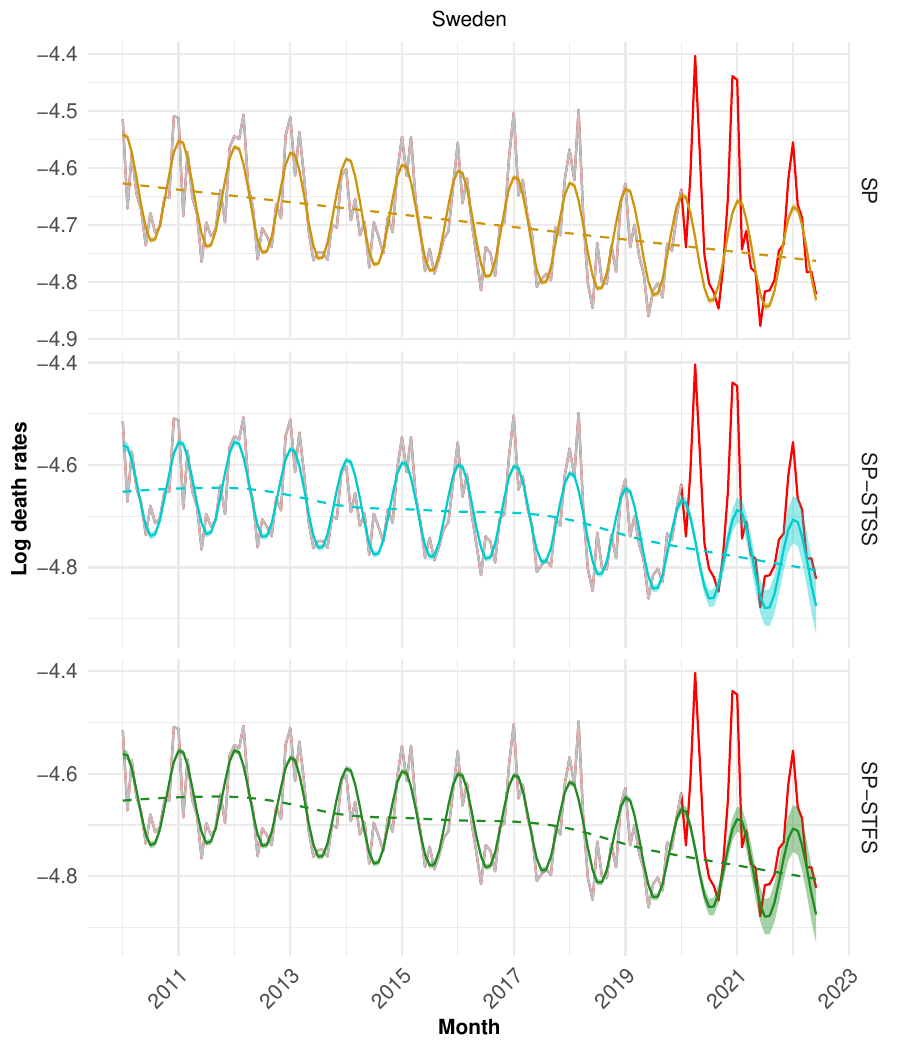}
		\caption{\textbf{Modelling and forecasting of CDRs (logarithmic scale) for Sweden with the SP, SP-STSS and SP-STFS models}. Trend function (dashed coloured line), fitted values (solid coloured line).}
		\label{fig:fig_fit_for}
	\end{center}
\end{figure}

Figure \ref{fig:fig_fit_for} shows the fit of the three models on the CDRs in the logarithmic scale for Sweden (and also shows the forecasts, which will be discussed later in Section \ref{section5}). The trend function $\boldsymbol{v}$ (dashed coloured line) and the fitted values $\boldsymbol{\hat{\mu}}$ (solid coloured line) are overlaid on the observed monthly CDRs (grey line). \\ 
Sweden has a declining CDR, according to the three models. However, the trend estimated is linear for the SP model and non linear for the SP-STSS and SP-STFS models, with the slope of the regression line changing over time. Furthermore, the SP-STSS and SP-STFS models better fit the mortality level during the summer, especially in the final years of the fitting period before COVID-19. The difference between the SP-STSS model and the SP-STFS model is that they assume, respectively, a varying seasonality and a fixed seasonality. The fitted models SP-STSS and SP-STFS did not capture a noticeable change in the seasonality. \\
Our approach is similar to that of \cite{eilers2008modulation} because we are not interested in catching the winter spikes. We do not want our model to overfit every year but, rather, to estimate an average pattern. A small excess mortality can happen every year between December and March due to variants of the influenza virus or a reduction in the effectiveness of the vaccines. We are interested in the counterfactual forecast and the excess mortality during a shock, which can be larger than the seasonal excess mortality and occur outside the usual period from December to March. Our approach differs from that of \cite{eilers2008modulation}, which ignores the months from December to March by giving them zero weights. We keep these months in our fitting period to use the information on the location of the winter peak.

\begin{figure}[h!t]
	\begin{center}
		\includegraphics[width=15cm,keepaspectratio]{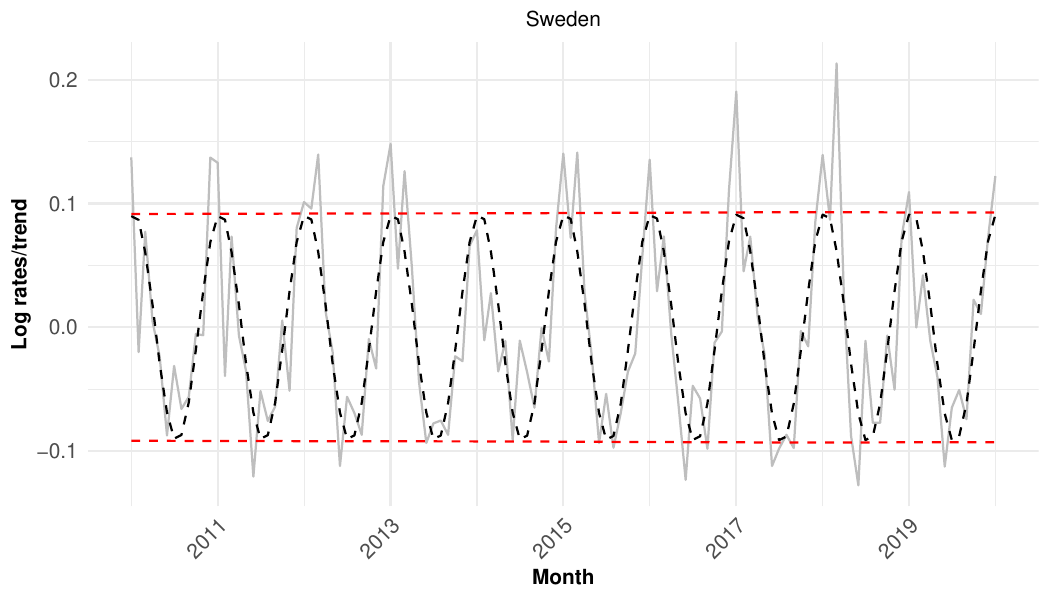}
		\caption{\textbf{Changes in the seasonal component over time for Sweden}. The seasonal component is estimated using the SP-STSS model. Detrended series (grey), modulated component (dashed black), and amplitude (dashed red).}
		\label{fig:amplitude_men}
	\end{center}
\end{figure}
As described by \cite{eilers2008modulation}, the amplitude of the seasonal variation, permits further inspection and measurement of the changes of the seasonal component over time. Figure \ref{fig:amplitude_men} shows the amplitude in the seasonal component obtained with the SP-STSS model in Sweden from 2010 to 2019. The grey line represents the detrended series, i.e., the ratio $\boldsymbol{y/\hat{\mu}}$, the dashed black line is the modulated component given by $f_{t}cos(wt)+g_{t}sin(wt)$, and the red line its amplitude $\boldsymbol{\rho}=\boldsymbol{\sqrt{(f^2+g^2)}}$. The seasonal amplitude remains constant over time in Sweden. 
\section{Results on the forecasts}
\label{section5}
The models are applied to obtain out-of-sample predictions of the monthly mortality. First, we explain our choice of the penalty order (\ref{section5.1}). We then present the forecasting accuracy of the three models (SP, SP-STSS, SP-STFS) on historical periods in the absence of mortality shocks (\ref{section5.2}). Finally, we calculate the excess mortality during the COVID-19 pandemic (\ref{section5.3}).
\subsection{Choice of the order of the penalty}
\label{section5.1}
\begin{figure}[!b]
	\includegraphics[width=15cm,keepaspectratio]{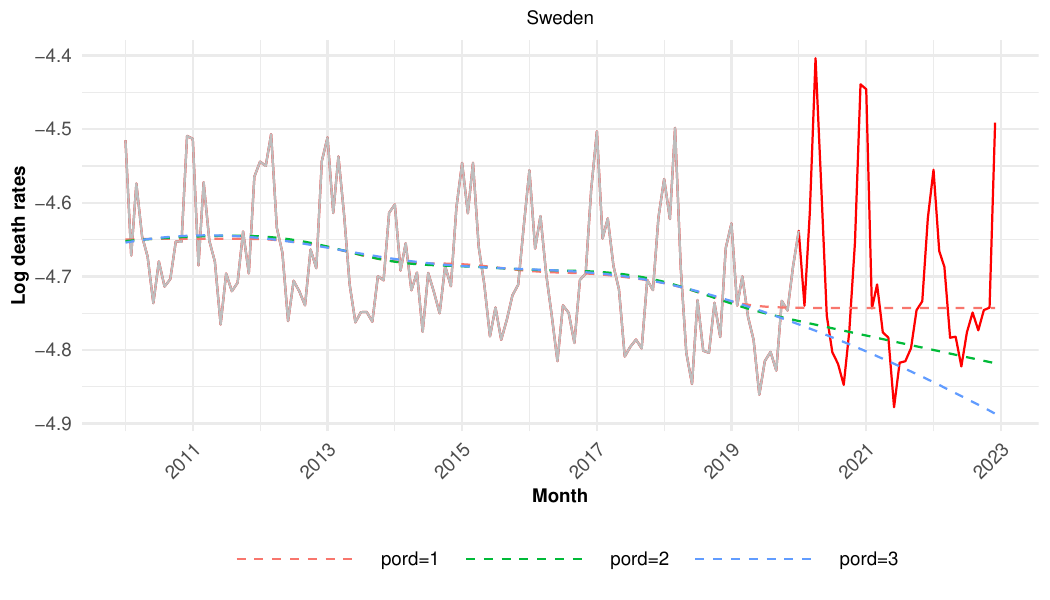}
	\caption{\textbf{Effect of the order of the penalty on the forecasts of the trend function (logarithmic scale) in Sweden}. The trend function is estimated with the model SP-STSS.}
	\label{fig:fig_penalty}
\end{figure}

The forecasting method works by extrapolating the regression coefficients, and the penalty on the coefficients ensures their smoothness. The order of the penalty determines whether the coefficients are approximately constant (order 1), linear (order 2), or quadratic (order 3). As a consequence, the smoothness of the forecasts also depends on the order of the penalty. Figure \ref{fig:fig_penalty} shows the effect of changing the penalty's order on the trend predictions. The second order for the trend, i.e., a linear extrapolation, best approximates the mortality trend, while the first order penalty for the modulation functions gives more realistic predictions and $95\%$ PIs.

Out-of-sample validation was used to choose the penalty order that provides better forecasts. We compared the following penalties: first order for the trend and seasonality, second order for both the trend and seasonality, first order for the trend and second order for the seasonality, and second order for the trend and first order for the seasonality. The results are shown in Table \ref{tab:best1} for the death counts and Table \ref{tab:best2} for the CDRs (\ref{appendixB}). The model with a second-order penalty for the trend and a first-order penalty for the seasonality provided better forecasting accuracy and was then chosen.

The penalty hyper-parameters $\lambda_{1}$ for the trend, and $\lambda_{2}$ for the seasonal component were chosen through a rolling window on the estimation set. We used values from $10^4$ to $10^7$ to compute the predictions one year ahead and chose the hyper-parameters based on the minimum mean MAPE. We decided to use the MAPE minimisation instead of the BIC minimisation, as our analysis focuses on forecasting (rather than modelling). This procedure was applied to both the SP-STSS model and the SP-STFS model.
\subsection{Evaluation of forecasting accuracy}
\label{section5.2}
We evaluated the forecasting accuracy of the three models (SP, SP-STSS, and SP-STFS) via out-of-sample validation. A rolling window of 5 years and a rolling window of 10 years were used to predict the CDRs one year ahead. The accuracy of the forecasts is measured based on the RMSE and MAPE. As in the case of the BIC comparisons, the series of CDRs allow for a sufficient number of years to evaluate the forecasts, for both the 5-year rolling window and the 10-year rolling window.

Table \ref{tab:mape} shows how well each model could have predicted the CDRs in the period before the COVID-19 pandemic. When using 5 years as a fitting period, the preferred model is the SP for the majority of the countries. However,  when using 10 years to fit the models, the forecasting accuracy improves, and the SP-STFS model becomes the preferred one. According to the mean MAPE (and mean RMSE), the SP-STFS would have been more accurate in predicting mortality one year ahead in 13 cases out of 25 (and 13 cases for the RMSE) than either the SP model (7 and 9 cases) or the SP-STSS model (4 and 3 cases). Therefore, for those countries, the trend is better approximated by more flexible models, rather than a linear interpolation.

Figure \ref{fig:fig_fit_for} plots the model-specific forecasts of the trend function $\boldsymbol{v}$ and the predicted values $\boldsymbol{\hat{\mu}}$ using a second-order penalty for the trend and a first-order penalty for the modulation functions. The observed monthly CDRs $\boldsymbol{y}$ are overlaid in grey on the fitting period (2011 to February 2020) and in red on the forecasting period (February 2020 to June 2022). The forecast of the trend is decreasing according to the three models, and the PIs are generally wider for the SP-STSS and SP-STFS models than they are for the SP model. The forecast from March 2020 to December 2022 can be considered as counterfactual mortality had the health shock not occurred. The difference between the forecasts and the observed mortality represents an estimate of the excess mortality attributable to the health shock. Depending on the model, the baseline mortality might differ and lead to different estimations of the excess mortality.

In an epidemic year, a significant excess mortality (i.e., outside the PI) is usually recorded in the months between December and March, when the winter peak occurs. In these months, mortality might exceed the expected mortality with an intensity that depends on the severity of the seasonal influenza. The excess mortality recorded during the COVID-19 pandemic from March 2020 until 2022 is visibly larger than in the preceding years. Furthermore, the excess in 2020, corresponding to the first COVID-19 wave, occurred outside the normal epidemic period. After the first wave, the excess mortality throughout the pandemic years 2020-21 and 2021-22 reaches a lower peak and follows the seasonal behavior of influenza deaths.  

\begin{table}[t]
\begingroup\fontsize{8}{10}\selectfont
\begin{longtable}[c]{>{}l|cc>{}c|cc>{}c|cc>{}c|clc}
	\caption{\label{tab:mape}\textbf{Mean RMSE and MAPE on one-year ahead CDRs (x 1000) in 25 European countries for multiple fitting periods based on a rolling-window scheme}. Three models (SP, SP-STSS, and SP-STFS) and two lengths of the fitting period are compared (5 and 10 years).}\\
	\toprule
	\multicolumn{1}{c}{ } & \multicolumn{6}{c}{5 years series} & \multicolumn{6}{c}{10 years series} \\
	\cmidrule(l{3pt}r{3pt}){2-7} \cmidrule(l{3pt}r{3pt}){8-13}
	\multicolumn{1}{c}{ } & \multicolumn{3}{c}{RMSE} & \multicolumn{3}{c}{MAPE} & \multicolumn{3}{c}{RMSE} & \multicolumn{3}{c}{MAPE} \\
	\cmidrule(l{3pt}r{3pt}){2-4} \cmidrule(l{3pt}r{3pt}){5-7} \cmidrule(l{3pt}r{3pt}){8-10} \cmidrule(l{3pt}r{3pt}){11-13}
	Country & SP & STSS & STFS & SP & STSS & STFS & SP & STSS & STFS & SP & STSS & STFS\\
	\midrule
	Austria & 0.52 & 0.53 & 0.53 & 4.08 & 4.12 & 4.06 & 0.51 & 0.51 & 0.5 & 4.1 & 3.93 & 3.9\\
	Bulgaria & 0.98 & 1.01 & 1.01 & 5 & 5.24 & 5.19 & 0.98 & 0.96 & 0.97 & 4.85 & 4.8 & 4.83\\
	Croatia & 0.8 & 0.83 & 0.82 & 4.93 & 5.15 & 5.06 & 0.77 & 0.78 & 0.78 & 4.24 & 4.41 & 4.43\\
	Czechia & 0.56 & 0.56 & 0.55 & 3.89 & 3.82 & 3.68 & 0.54 & 0.53 & 0.52 & 3.71 & 3.63 & 3.59\\
	Denmark & 0.5 & 0.51 & 0.5 & 3.85 & 3.88 & 3.85 & 0.48 & 0.48 & 0.47 & 3.76 & 3.64 & 3.56\\
	\addlinespace
	Estonia & 0.71 & 0.72 & 0.72 & 4.39 & 4.47 & 4.46 & 0.69 & 0.65 & 0.65 & 4.33 & 4.13 & 4.07\\
	Finland & 0.49 & 0.5 & 0.5 & 3.82 & 3.9 & 3.93 & 0.47 & 0.47 & 0.46 & 3.59 & 3.59 & 3.56\\
	France & 0.52 & 0.54 & 0.53 & 4.13 & 4.46 & 4.38 & 0.47 & 0.49 & 0.47 & 3.81 & 4.09 & 3.85\\
	Germany & 0.59 & 0.61 & 0.59 & 3.84 & 4.06 & 3.93 & 0.6 & 0.59 & 0.58 & 3.91 & 3.69 & 3.62\\
	Greece & 0.76 & 0.76 & 0.75 & 5.38 & 5.35 & 5.28 & 0.74 & 0.74 & 0.75 & 5.12 & 5.26 & 5.27\\
	\addlinespace
	Hungary & 0.8 & 0.83 & 0.82 & 4.3 & 4.58 & 4.46 & 0.74 & 0.74 & 0.74 & 3.8 & 3.95 & 3.94\\
	Iceland & 0.65 & 0.65 & 0.65 & 8.07 & 8.09 & 8.09 & 0.59 & 0.59 & 0.59 & 7.62 & 7.59 & 7.59\\
	Ireland & 0.49 & 0.48 & 0.49 & 5.36 & 5.38 & 5.46 & 0.46 & 0.4 & 0.4 & 5.48 & 4.86 & 4.86\\
	Italy & 0.71 & 0.78 & 0.77 & 5.33 & 5.88 & 5.84 & 0.68 & 0.73 & 0.72 & 4.91 & 5.35 & 5.3\\
	Lithuania & 0.84 & 0.82 & 0.81 & 5.2 & 4.99 & 4.96 & 1.02 & 0.87 & 0.87 & 6.2 & 5.21 & 5.22\\
	\addlinespace
	Luxembourg & 0.63 & 0.64 & 0.64 & 6.84 & 7.03 & 7.03 & 0.57 & 0.57 & 0.57 & 6.24 & 6.3 & 6.3\\
	Netherlands & 0.47 & 0.48 & 0.48 & 3.98 & 4.2 & 4.17 & 0.49 & 0.46 & 0.45 & 4.25 & 3.9 & 3.85\\
	Norway & 0.52 & 0.5 & 0.51 & 4.82 & 4.68 & 4.82 & 0.49 & 0.46 & 0.49 & 4.96 & 4.49 & 4.97\\
	Poland & 0.53 & 0.57 & 0.55 & 3.76 & 4.03 & 3.97 & 0.56 & 0.54 & 0.54 & 3.96 & 3.71 & 3.71\\
	Portugal & 0.84 & 0.89 & 0.88 & 5.79 & 6.31 & 6.32 & 0.82 & 0.84 & 0.83 & 5.54 & 5.82 & 5.77\\
	\addlinespace
	Romania & 0.83 & 0.85 & 0.85 & 5.24 & 5.36 & 5.35 & 0.73 & 0.76 & 0.78 & 4.24 & 4.68 & 4.72\\
	Slovenia & 0.6 & 0.62 & 0.61 & 4.89 & 4.93 & 4.93 & 0.58 & 0.58 & 0.58 & 4.35 & 4.36 & 4.32\\
	Spain & 0.66 & 0.69 & 0.68 & 5.53 & 5.67 & 5.56 & 0.64 & 0.62 & 0.62 & 5.38 & 5 & 4.99\\
	Sweden & 0.51 & 0.51 & 0.5 & 3.85 & 3.87 & 3.8 & 0.45 & 0.46 & 0.45 & 3.47 & 3.52 & 3.5\\
	Switzerland & 0.45 & 0.46 & 0.46 & 4.01 & 4.16 & 4.14 & 0.4 & 0.4 & 0.4 & 3.58 & 3.57 & 3.5\\
	\addlinespace
	No. countries * & 17 & 2 & 6 & 18 & 1 & 6 & 7 & 4 & 14 & 9 & 3 & 13\\
	\bottomrule
	\multicolumn{13}{l}{\rule{0pt}{1em}\textsuperscript{*} Number of countries in which the model performs the best}\\
\end{longtable}
\endgroup{}
\end{table}
\subsection{Excess mortality during the COVID-19 pandemic}
\label{section5.3}
\begin{figure}[h!p]
	\includegraphics[width=14cm,keepaspectratio]{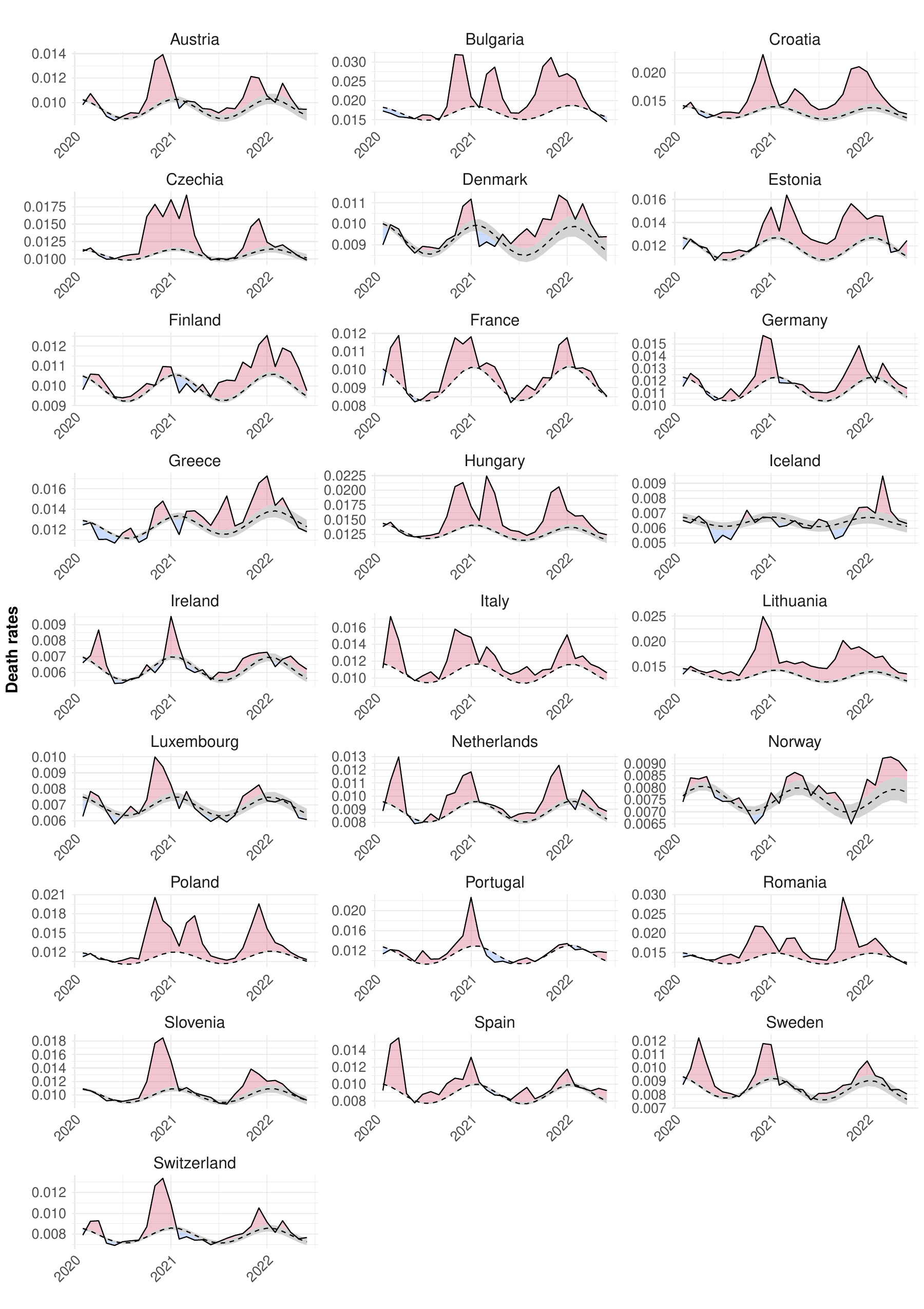}
	\caption{\textbf{Monthly excess CDRs from February 2020 through June 2022 in 25 European countries}. Recorded CDRs (solid line), predictions of the expected CDRs (dashed line), 95\% prediction intervals around the mean prediction (grey shaded area), excess CDRs (red shaded area), and deficit CDRs (blue shaded area).}
	\label{fig:excess_deaths_EU}
\end{figure}

Figure \ref{fig:excess_deaths_EU} shows the observed CDRs (continuous line) and the expected CDRs (dashed line) together with the 95\% PIs (grey shaded area) from March 2020 to June 2022 for the 25 analysed European countries. The expected CDRs and the 95\% PIs are estimated on a common period (2010 to January 2020) using the SP-STFS model. Only for Ireland, Lithuania, Norway, and Poland, a SP-STSS model was used based on accuracy analysis (Table \ref{tab:mape}). The prediction intervals represent the uncertainty in the CDRs. The CDRs in excess or deficit were obtained by subtracting the expected CDRs, had the COVID-19 pandemic not occurred, from the observed CDRs. The excess and deficit mortality outside of the PIs is shaded, respectively, in red and blue in the plot. The PIs conveniently widen, when moving away from the starting point of the forecasts, indicating that the uncertainty increases with the forecasting horizon. Table \ref{tab:excess_rates_EU} reports the excess CDRs together with their 95\% credible intervals. Among the countries analysed, France, Ireland, Italy, the Netherlands, Spain, Sweden, and Switzerland were hit in the first wave and had a significant excess mortality. The remaining countries either saved lives (Bulgaria, Denmark, Estonia, Germany, Greece, Hungary, Iceland, Luxembourg, Poland, Romania) or did not show any significant change (Austria, Croatia, Czechia, Finland, Norway, Portugal, Slovenia). In the following two pandemic years 2020-21 and 2021-22, France, Ireland, Italy, Lithuania, the Netherlands, Spain, Sweden, and Switzerland still had a significant excess mortality. Some countries that avoided excess mortality in the first wave were hit later on: Austria, Bulgaria, Croatia, Czechia, Estonia, Germany, Greece, Hungary, Luxembourg, Poland, Portugal, Romania, and Slovenia. Denmark and Finland were hit only in the pandemic year 2021-22.

We also analyse the demographic distribution of the excess in four countries: Denmark, Italy, Spain, and Sweden. Table \ref{tab:excess_rates_AS} reports the excess ASDRs together with their 95\% credible intervals from March 2020 to June 2021 by sex and age groups. For ease of interpretation, we distinguished three pandemic phases: 1) the first wave (from March 2020 to June 2020), 2) the pandemic year 2020-21 (from July 2020 to June 2021), and 3) the pandemic year 2021-22 (from July 2021 to June 2022). In Denmark, neither of the two sexes experienced an excess in the ASDRs. Lives were saved during the first wave in men aged above 75 and in both pandemic periods in women aged above 85. In Italy and Spain, significant excess mortality occurred during the first wave and during the pandemic year 2020-21. The excess occurred in all age groups, but most of the excess was concentrated in the oldest age groups. A male disadvantage at all ages occurred in Italy and Spain in the two pandemic periods. In Sweden, an excess in the ASDRs was registered during the first wave, with a male disadvantage at older ages. In the pandemic year 2020-2021, the excess in the ASDRs was found in males at all ages and females aged 75-84. Overall, during the two pandemic phases, the toll of the pandemic hit harder in those aged 75 years and older. A slight disadvantage was found between sexes in Denmark and Sweden, and a male disadvantage was largest in Italy and Spain.

\begin{table}[h!t]
\begingroup\fontsize{10}{12}\selectfont
\begin{longtable}[c]{ccccc}
	\caption{\label{tab:excess_rates_AS}\textbf{Excess ASDRs in Denmark, Italy, Spain, Sweden, by sex and age from March 2020 to June 2021}. The excess is divided into two pandemic periods: the first COVID-19 wave and the first pandemic year after the shock.}\\
	\toprule
	\multicolumn{1}{c}{ } & \multicolumn{4}{c}{Excess ASDRs (prediction  intervals)} \\
	\cmidrule(l{3pt}r{3pt}){2-5}
	Age & Denmark & Italy & Spain & Sweden\\
	\addlinespace[0.3em]
	\midrule
	\multicolumn{5}{l}{\textbf{Men}}\\
	\midrule
	\addlinespace[0.3em]
	\multicolumn{5}{l}{\textbf{March to June 2020}}\\
	\hspace{1em}\hspace{1em}0-64 & -0.6 (-1 ; -0.2) & 1.1 (1 ; 1.2) & 0.7 (0.6 ; 0.8) & 0.8 (0.7 ; 1)\\
	\hspace{1em}\hspace{1em}65-74 & -2.8 (-6.3 ; 0.7) & 18.6 (17.6 ; 19.6) & 17.8 (16.2 ; 19.4) & 9.8 (8.4 ; 11.3)\\
	\hspace{1em}\hspace{1em}75-84 & -7.5 (-13.6 ; -1.5) & 56.2 (54.1 ; 58.1) & 65.5 (61.9 ; 69) & 36 (33 ; 39)\\
	\hspace{1em}\hspace{1em}85 & -34.4 (-51.9 ; -17.1) & 113.3 (107.8 ; 118.8) & 205.3 (194.9 ; 215.5) & 139.9 (124.4 ; 155.2)\\
	\addlinespace[0.3em]
	\multicolumn{5}{l}{\textbf{July 2020 to June 2021}}\\
	\hspace{1em}\hspace{1em}0-64 & -0.6 (-1.9 ; 0.6) & 3.7 (3.4 ; 3.9) & 1.9 (1.6 ; 2.1) & 0.9 (0.5 ; 1.3)\\
	\hspace{1em}\hspace{1em}65-74 & -7 (-19.2 ; 4.6) & 51.1 (47.6 ; 54.5) & 24.4 (17.9 ; 30.8) & 9.7 (5.9 ; 13.5)\\
	\hspace{1em}\hspace{1em}75-84 & 4.6 (-13.7 ; 22.4) & 120.7 (113.9 ; 127.4) & 78.7 (63.6 ; 93.3) & 40.4 (32.5 ; 48.2)\\
	\hspace{1em}\hspace{1em}85 & -2.4 (-47.3 ; 41.5) & 255.7 (238.8 ; 272.3) & 209.3 (163.6 ; 253.7) & 72.4 (23.6 ; 119.9)\\
	\addlinespace[0.3em]
	\midrule
	\multicolumn{5}{l}{\textbf{Women}}\\
	\midrule
	\addlinespace[0.3em]
	\multicolumn{5}{l}{\textbf{March to June 2020}}\\
	\hspace{1em}\hspace{1em}0-64 & -0.1 (-0.3 ; 0.1) & 0.2 (0.2 ; 0.3) & 0.5 (0.4 ; 0.6) & 0.1 (0 ; 0.2)\\
	\hspace{1em}\hspace{1em}65-74 & -2.5 (-5 ; 0) & 6 (5.3 ; 6.7) & 6.9 (6.4 ; 7.4) & 0.3 (-0.9 ; 1.5)\\
	\hspace{1em}\hspace{1em}75-84 & -2.5 (-8.2 ; 3.2) & 25.2 (24.2 ; 26.3) & 37.8 (35.6 ; 39.9) & 22.9 (19.1 ; 26.6)\\
	\hspace{1em}\hspace{1em}85 & -52.5 (-64.5 ; -40.7) & 106.3 (102.7 ; 109.8) & 165.8 (158.6 ; 173) & 103.7 (89.1 ; 118.1)\\
	\addlinespace[0.3em]
	\multicolumn{5}{l}{\textbf{July 2020 to June 2021}}\\
	\hspace{1em}\hspace{1em}0-64 & 0.3 (-0.1 ; 0.8) & 1.4 (1.2 ; 1.5) & 0.5 (0.3 ; 0.7) & -0.1 (-0.4 ; 0.2)\\
	\hspace{1em}\hspace{1em}65-74 & -0.3 (-8.9 ; 7.8) & 21.7 (19 ; 24.3) & 9.7 (8.2 ; 11.1) & -2.1 (-5.2 ; 0.9)\\
	\hspace{1em}\hspace{1em}75-84 & -0.9 (-20.5 ; 17.8) & 62.3 (59.2 ; 65.3) & 48.4 (39.5 ; 57.1) & 31.1 (18.5 ; 43.3)\\
	\hspace{1em}\hspace{1em}85 & -41.2 (-72.3 ; -10.7) & 179.1 (168.5 ; 189.7) & 103.1 (68.9 ; 136.5) & 50.3 (-5.8 ; 104.4)\\
	\bottomrule
\end{longtable}
\endgroup{}
\end{table}

\begin{table}[h!t]
\end{table}
\section{Discussion}
Before the COVID-19 pandemic, short-term mortality forecasting during an epidemiological year was mainly studied in the domain of epidemiology. After the COVID-19 pandemic, a collective effort to provide data and evidence on the health shock increased interest and so an increase in the number of publications on excess mortality from different fields was seen. Numerous estimates of excess mortality have been published using various approaches to forecast the expected mortality without a shock \citep{kontis2020magnitude, islam2021excess}. Among the studies using regression models, many used a Serfling-Poisson regression \citep{serfling1963methods, euromomo2017european} or more flexible versions of the model that assumed smoothed effects \citep{aburto2021estimating, scortichini2020excess}. However, comparisons of these approaches are lacking. Few studies have investigated the sensitivity of excess mortality estimates, but either did not study the accuracy in forecasting the expected mortality on historical periods \citep{nepomuceno2022sensitivity} or did not consider smoothed effects for both the trend and seasonal effects \citep{scholey2021robustness}. Our current study considers the same family of models and estimation procedure. We allow for different degrees of flexibility of the Serfling-Poisson model with: 1) parametric effect for the trend and seasonality, 2) non-parametric effect for the trend and parametric effect for the seasonality, and 3) non-parametric effect for the trend and seasonality.

For all-cause mortality, the smoothness of the trend appeared to be a desirable feature when forecasting in the short term. The smoothness of the seasonal component might be a useful assumption in other settings, for instance, when the data show a trend of decreasing or increasing seasonal amplitude. Progress in mortality could lead to a decreasing seasonality, which might be better captured with a smooth seasonal component. This could be the case, for instance, for mortality declines in certain historical periods \citep{ledberg2020large} or for specific causes of death, such as infectious diseases \citep{schluter2020long} or cardiovascular diseases \citep{crawford2003changes}.

We illustrated how the SP-STFS model could be used to forecast mortality during the COVID-19 pandemic and analysed the vulnerability of a population during the health shock. In order to compare the mortality experiences during the pandemic, one needs to account for the different population sizes and age structures between countries. This can be achieved by considering the exposure in each age group and modelling the mortality rates. All models permit the inclusion of the exposures as an offset term. Furthermore, the SP-STFS model showed a competitive forecasting accuracy for both death counts and death rates. Therefore, one could also perform demographic comparisons using the SP-STFS model on age-specific or age-standardized death rates.

One limitation of the SP-STFS and SP-STSS models for forecasting mortality is that the PIs of the forecasts increase rather quickly. Therefore, we advise their use for one to three years in the future to predict the baseline mortality and to detect excess mortality. One year is usually used to monitor the severity of the influenza season, while two to three years is the time window for infectious disease-related pandemics to extinguish themselves.

Further work on short-term forecasting using the SP-STFS model could extend the model to forecast all age groups simultaneously. For instance, a bidimensional model could simultaneously forecast the age and year directions \citep{eilers2008modulation, currie2004smoothing}. Furthermore, modelling the median instead of the mean could better capture the asymmetric shape of the seasonality, because mortality shows a percentage increase during the winter that is greater than the percentage decrease during the summer.
\section{Conclusion}
The aim of this paper is to propose a new methodology to model and forecast baseline mortality that is more flexible than the widely used Serfling-Poisson model. Specifically, we compare the forecast of the Serfling-Poisson model with two versions of the model, assuming either fixed or smooth seasonal components. The results show that the model with smooth trend and seasonal component (SP-STSS) better fits the historical time series of CDRs, followed by the model with smooth trend and fixed seasonal component (SP-STFS). However, the model with smooth trend and fixed seasonal component (SP-STFS) produces more accurate predictions of the expected CDRs, indicating that a model with a smooth seasonal component may overfit the data. Therefore, accounting for demographic changes by considering smooth trends over longer reference periods and across age groups is essential for reliable estimates of expected mortality. Our short-term mortality predictions come with prediction intervals, whose widths indicate the level of uncertainty associated with the forecast.
\section*{Author contribution}
A.-E.L. contributed to design the study (Methodology), conducted the data analysis (Data curation and Formal analysis), visualised the results (Visualization) and drafted the manuscript (Writing- Original draft preparation). S.R. conceived the research idea (Conceptualization). S.R. and U.B. critically contributed to design the study (Methodology), commented and reviewed the manuscript (Writing- Reviewing and Editing). All authors read and approved the final version of the manuscript.
\section*{Funding information}
This work was supported by the AXA Research Fund’s AXA Chair in Longevity Research (A.-E.L.) and the SCOR Foundation’s SCOR Chair on Mortality Research 2023-2026 (S.R.).
\section*{Competing interests}
The authors declare that they have no competing interests.
\section*{Code availability}
The codes are publicly available at the online link \url{https://github.com/AinhoaLeger/Shortcasting-Seasonal-Mortality}.
\bibliographystyle{elsarticle-harv}
\bibliography{references}
\newpage
\appendix
\section{Model performance}
\label{appendixA}
Table \ref{tab:bic2} shows the mean BIC values obtained by fitting the three models (SP, SP-STSS, and SP-STFS) on the death counts from 2010 to 2019 on two rolling windows of 5 years and 10 years. The mean BIC favours the model SP-STSS in almost all the countries (21 countries). Either using a 5-year or a 10-year rolling window, the model SP-STSS better fits the series than the SP and the SP-STFS model.
\begin{table}[h!]
\begingroup\fontsize{8}{10}\selectfont
\begin{longtable}[c]{>{}l|cc>{}c|ccc}
	\caption{\label{tab:bic2}\textbf{Mean BIC on death counts in 25 European countries for multiple fitting periods based on a rolling-window scheme}. Three models (SP, SP-STSS, and SP-STFS) and two lengths of the time series are compared (5 and 10 years).}\\
	\toprule
	\multicolumn{1}{c}{ } & \multicolumn{3}{c}{5 years series} & \multicolumn{3}{c}{10 years series} \\
	\cmidrule(l{3pt}r{3pt}){2-4} \cmidrule(l{3pt}r{3pt}){5-7}
	Country & SP & SP-STSS & SP-STFS & SP & SP-STSS & SP-STFS\\
	\midrule
	Austria & 1,040 & 994 & 1,021 & 2,047 & 1,996 & 2,019\\
	Bulgaria & 1,885 & 1,828 & 1,839 & 3,762 & 3,662 & 3,708\\
	Croatia & 838 & 796 & 817 & 1,641 & 1,605 & 1,635\\
	Czechia & 1,325 & 1,273 & 1,297 & 2,510 & 2,351 & 2,382\\
	Denmark & 650 & 643 & 654 & 1,357 & 1,310 & 1,332\\
	\addlinespace
	Estonia & 241 & 250 & 251 & 484 & 485 & 487\\
	Finland & 598 & 597 & 609 & 1,219 & 1,205 & 1,219\\
	France & 8,158 & 7,492 & 7,941 & 17,074 & 16,109 & 16,605\\
	Germany & 10,617 & 9,616 & 10,153 & 21,838 & 20,120 & 20,961\\
	Greece & 2,644 & 2,511 & 2,576 & 5,034 & 4,864 & 4,945\\
	\addlinespace
	Hungary & 2,162 & 2,056 & 2,124 & 4,140 & 4,048 & 4,083\\
	Iceland & 96 & 109 & 109 & 201 & 217 & 216\\
	Ireland & 698 & 691 & 706 & 1,495 & 1,411 & 1,464\\
	Italy & 11,279 & 10,048 & 10,592 & 22,338 & 20,633 & 21,466\\
	Lithuania & 596 & 583 & 592 & 1,452 & 1,235 & 1,246\\
	\addlinespace
	Luxembourg & 109 & 120 & 120 & 219 & 234 & 233\\
	Netherlands & 1,862 & 1,781 & 1,820 & 3,804 & 3,610 & 3,627\\
	Norway & 621 & 595 & 629 & 1,262 & 1,178 & 1,244\\
	Poland & 4,861 & 4,488 & 4,687 & 9,811 & 8,967 & 9,451\\
	Portugal & 3,535 & 3,400 & 3,496 & 6,805 & 6,705 & 6,763\\
	\addlinespace
	Romania & 4,165 & 3,778 & 3,881 & 8,973 & 8,142 & 8,614\\
	Slovenia & 315 & 316 & 323 & 638 & 641 & 650\\
	Spain & 9,554 & 8,710 & 9,125 & 19,192 & 18,237 & 18,792\\
	Sweden & 1,100 & 1,074 & 1,101 & 2,225 & 2,214 & 2,234\\
	Switzerland & 826 & 795 & 826 & 1,634 & 1,613 & 1,628\\
	\addlinespace
	No. countries * & 4 & 21 & 0 & 4 & 21 & 0\\
	\bottomrule
	\multicolumn{7}{l}{\rule{0pt}{1em}\textsuperscript{*} Number of countries in which the model performs the best}\\
\end{longtable}
\endgroup{}
\end{table}
\newpage
\section{Best model based on P-splines}
\label{appendixB}
The RMSE and MAPE of the predictions on the death counts and CDRs are shown in Table \ref{tab:best1} Table \ref{tab:best2}, respectively, for 25 European countries. A 5-year fitting period is used. Four combinations of the order of the penalties are considered: (1) order 1 for the trend and the seasonality, (2) order 2 for the trend and the seasonality, (3) order 1 for the trend and 2 for the seasonality, (4) order 2 for the trend and order 1 for the seasonality.

\begin{table}[h!t]
\begingroup\fontsize{7.5}{9.5}\selectfont
\begin{longtable}[c]{>{}l|ccc>{}c|ccc>{}c|ccl>{}c|cccc}
	\caption{\label{tab:best1}\textbf{Effect of choice of the order of the penalty on the forecasts of the death counts}. Mean RMSE and MAPE on death counts in 25 European countries for multiple fitting periods based on a rolling-window scheme. Four combinations of penalties  are considered: (1) order 1 on trend and seasonality, (2) order 2 on the trend and seasonality, (3) order 1 on the trend and 2 on the seasonality, (4) order 2 on the trend and order 1 on the seasonality.}\\
	\toprule
	\multicolumn{1}{c}{ } & \multicolumn{8}{c}{5 years series} & \multicolumn{8}{c}{10 years series} \\
	\cmidrule(l{3pt}r{3pt}){2-9} \cmidrule(l{3pt}r{3pt}){10-17}
	\multicolumn{1}{c}{ } & \multicolumn{4}{c}{RMSE} & \multicolumn{4}{c}{MAPE} & \multicolumn{4}{c}{RMSE} & \multicolumn{4}{c}{MAPE} \\
	\cmidrule(l{3pt}r{3pt}){2-5} \cmidrule(l{3pt}r{3pt}){6-9} \cmidrule(l{3pt}r{3pt}){10-13} \cmidrule(l{3pt}r{3pt}){14-17}
	Country & (1) & (2) & (3) & (4) & (1) & (2) & (3) & (4) & (1) & (2) & (3) & (4) & (1) & (2) & (3) & (4)\\
	\midrule
	Austria & 381 & 396 & 400 & 375 & 4.2 & 4.5 & 4.5 & 4.1 & 400 & 377 & 410 & 358 & 4.2 & 4.1 & 4.4 & 3.8\\
	Bulgaria & 591 & 642 & 616 & 619 & 4.8 & 5.3 & 5.1 & 5.2 & 551 & 574 & 560 & 561 & 4.7 & 4.9 & 4.9 & 4.7\\
	Croatia & 276 & 312 & 291 & 295 & 4.6 & 5.5 & 4.9 & 5.2 & 262 & 280 & 274 & 269 & 4.1 & 4.6 & 4.4 & 4.4\\
	Czechia & 485 & 536 & 530 & 489 & 3.8 & 4.4 & 4.3 & 3.8 & 467 & 501 & 496 & 469 & 3.7 & 4 & 3.9 & 3.7\\
	Denmark & 240 & 252 & 254 & 235 & 4.1 & 4.2 & 4.4 & 3.9 & 255 & 239 & 266 & 225 & 4.7 & 3.9 & 4.9 & 3.7\\
	\addlinespace
	Estonia & 89 & 84 & 93 & 80 & 5.4 & 4.7 & 5.6 & 4.4 & 102 & 74 & 105 & 73 & 6.6 & 4.3 & 6.7 & 4.2\\
	Finland & 232 & 240 & 247 & 223 & 4 & 4.3 & 4.4 & 3.9 & 247 & 222 & 255 & 211 & 4.2 & 3.9 & 4.5 & 3.6\\
	France & 2812 & 3163 & 3111 & 2906 & 4.2 & 4.9 & 4.8 & 4.4 & 2796 & 2877 & 2971 & 2642 & 4.2 & 4.4 & 4.5 & 4\\
	Germany & 4110 & 4815 & 4665 & 4213 & 4 & 4.9 & 4.7 & 4.1 & 4045 & 4529 & 4473 & 4033 & 3.8 & 4.3 & 4.4 & 3.7\\
	Greece & 681 & 735 & 721 & 688 & 5.1 & 5.8 & 5.5 & 5.4 & 716 & 697 & 730 & 674 & 5.4 & 5.5 & 5.5 & 5.3\\
	\addlinespace
	Hungary & 681 & 747 & 728 & 691 & 4.5 & 5.1 & 5 & 4.6 & 635 & 643 & 662 & 610 & 4.3 & 4.2 & 4.6 & 3.9\\
	Iceland & 17 & 18 & 18 & 17 & 8.5 & 8.6 & 8.8 & 8.1 & 19 & 17 & 19 & 16 & 9.1 & 8 & 9.3 & 7.7\\
	Ireland & 179 & 180 & 185 & 177 & 5.4 & 5.6 & 5.9 & 5.4 & 171 & 160 & 179 & 156 & 5.1 & 4.9 & 5.5 & 4.8\\
	Italy & 3500 & 4207 & 3896 & 3805 & 5.2 & 6.5 & 6 & 5.8 & 3421 & 3878 & 3655 & 3633 & 4.9 & 5.7 & 5.3 & 5.4\\
	Lithuania & 223 & 226 & 237 & 212 & 5.2 & 5.3 & 5.5 & 4.9 & 222 & 222 & 225 & 219 & 5.2 & 5.1 & 5.1 & 5.1\\
	\addlinespace
	Luxembourg & 26 & 29 & 28 & 27 & 6.6 & 7.4 & 6.9 & 7 & 28 & 26 & 28 & 26 & 6.9 & 6.5 & 6.9 & 6.3\\
	Netherlands & 678 & 702 & 709 & 670 & 4.2 & 4.5 & 4.6 & 4.2 & 718 & 682 & 744 & 643 & 4.4 & 4.2 & 4.7 & 3.9\\
	Norway & 207 & 214 & 219 & 203 & 4.7 & 4.9 & 5 & 4.7 & 196 & 197 & 201 & 191 & 4.7 & 4.5 & 4.7 & 4.5\\
	Poland & 1741 & 1986 & 1898 & 1805 & 3.9 & 4.6 & 4.3 & 4.1 & 1766 & 1862 & 1876 & 1724 & 3.8 & 4.1 & 4.2 & 3.7\\
	Portugal & 734 & 845 & 814 & 773 & 5.8 & 7 & 6.5 & 6.3 & 732 & 756 & 757 & 729 & 5.7 & 6 & 5.8 & 5.9\\
	\addlinespace
	Romania & 1343 & 1487 & 1399 & 1457 & 4.7 & 5.4 & 4.9 & 5.3 & 1254 & 1333 & 1308 & 1287 & 4.4 & 4.8 & 4.6 & 4.7\\
	Slovenia & 103 & 111 & 109 & 104 & 4.7 & 5.2 & 5 & 4.9 & 104 & 104 & 108 & 99 & 4.5 & 4.6 & 4.8 & 4.4\\
	Spain & 2455 & 2797 & 2673 & 2581 & 5.2 & 6.4 & 5.8 & 5.7 & 2413 & 2626 & 2597 & 2416 & 4.8 & 5.5 & 5.2 & 5\\
	Sweden & 394 & 424 & 424 & 395 & 3.8 & 4.2 & 4.2 & 3.9 & 369 & 395 & 395 & 368 & 3.7 & 3.9 & 4 & 3.5\\
	Switzerland & 298 & 332 & 329 & 302 & 4.1 & 4.7 & 4.7 & 4.2 & 286 & 286 & 299 & 272 & 3.9 & 3.8 & 4.1 & 3.6\\
	\addlinespace
	No. * & 16 & 0 & 0 & 9 & 15 & 0 & 0 & 10 & 6 & 0 & 0 & 19 & 7 & 0 & 1 & 17\\
	\bottomrule
	\multicolumn{17}{l}{\rule{0pt}{1em}\textsuperscript{*} Number of countries in which the model performs the best}\\
\end{longtable}
\endgroup{}
\end{table}

\begin{table}[h!t]
\begingroup\fontsize{7.5}{9.5}\selectfont
\begin{longtable}[c]{>{}l|ccc>{}c|ccc>{}c|ccl>{}c|cccc}
	\caption{\label{tab:best2}\textbf{Effect of choice of the order of the penalty on the forecasts of the CDRs}. Mean RMSE and MAPE on CDRs in 25 European countries for multiple fitting periods based on a rolling-window scheme. Four combinations of penalties  are considered: (1) order 1 on trend and seasonality, (2) order 2 on the trend and seasonality, (3) order 1 on the trend and 2 on the seasonality, (4) order 2 on the trend and order 1 on the seasonality.}\\
	\toprule
	\multicolumn{1}{c}{ } & \multicolumn{8}{c}{5 years series} & \multicolumn{8}{c}{10 years series} \\
	\cmidrule(l{3pt}r{3pt}){2-9} \cmidrule(l{3pt}r{3pt}){10-17}
	\multicolumn{1}{c}{ } & \multicolumn{4}{c}{RMSE} & \multicolumn{4}{c}{MAPE} & \multicolumn{4}{c}{RMSE} & \multicolumn{4}{c}{MAPE} \\
	\cmidrule(l{3pt}r{3pt}){2-5} \cmidrule(l{3pt}r{3pt}){6-9} \cmidrule(l{3pt}r{3pt}){10-13} \cmidrule(l{3pt}r{3pt}){14-17}
	Country & (1) & (2) & (3) & (4) & (1) & (2) & (3) & (4) & (1) & (2) & (3) & (4) & (1) & (2) & (3) & (4)\\
	\midrule
	Austria & 0.54 & 0.57 & 0.57 & 0.53 & 4.29 & 4.5 & 4.55 & 4.12 & 0.54 & 0.54 & 0.56 & 0.51 & 4.28 & 4.18 & 4.44 & 3.93\\
	Bulgaria & 0.97 & 1.05 & 1.01 & 1.01 & 4.77 & 5.37 & 5.02 & 5.24 & 1.05 & 0.98 & 1.07 & 0.96 & 5.1 & 4.99 & 5.24 & 4.8\\
	Croatia & 0.79 & 0.88 & 0.83 & 0.83 & 4.58 & 5.45 & 4.86 & 5.15 & 0.84 & 0.82 & 0.87 & 0.78 & 4.56 & 4.58 & 4.78 & 4.41\\
	Czechia & 0.55 & 0.62 & 0.61 & 0.56 & 3.78 & 4.34 & 4.26 & 3.82 & 0.52 & 0.56 & 0.56 & 0.53 & 3.6 & 3.91 & 3.88 & 3.63\\
	Denmark & 0.55 & 0.54 & 0.58 & 0.51 & 4.52 & 4.18 & 4.79 & 3.88 & 0.63 & 0.51 & 0.65 & 0.48 & 5.63 & 3.87 & 5.69 & 3.64\\
	\addlinespace
	Estonia & 0.75 & 0.75 & 0.79 & 0.72 & 4.93 & 4.74 & 5.22 & 4.47 & 0.79 & 0.67 & 0.81 & 0.65 & 5.46 & 4.26 & 5.63 & 4.13\\
	Finland & 0.5 & 0.54 & 0.54 & 0.5 & 3.89 & 4.24 & 4.29 & 3.9 & 0.5 & 0.49 & 0.52 & 0.47 & 3.88 & 3.85 & 4.09 & 3.59\\
	France & 0.53 & 0.59 & 0.59 & 0.54 & 4.4 & 4.87 & 5.04 & 4.46 & 0.49 & 0.53 & 0.54 & 0.49 & 4.09 & 4.57 & 4.56 & 4.09\\
	Germany & 0.6 & 0.7 & 0.69 & 0.61 & 4.05 & 4.8 & 4.79 & 4.06 & 0.6 & 0.66 & 0.66 & 0.59 & 3.85 & 4.27 & 4.44 & 3.69\\
	Greece & 0.75 & 0.81 & 0.8 & 0.76 & 5.18 & 5.77 & 5.57 & 5.35 & 0.79 & 0.77 & 0.81 & 0.74 & 5.44 & 5.52 & 5.64 & 5.26\\
	\addlinespace
	Hungary & 0.8 & 0.9 & 0.86 & 0.83 & 4.34 & 5.03 & 4.81 & 4.58 & 0.75 & 0.78 & 0.78 & 0.74 & 4.04 & 4.24 & 4.31 & 3.95\\
	Iceland & 0.64 & 0.68 & 0.68 & 0.65 & 8.54 & 8.53 & 8.87 & 8.09 & 0.59 & 0.61 & 0.61 & 0.59 & 7.66 & 7.86 & 8 & 7.59\\
	Ireland & 0.57 & 0.49 & 0.58 & 0.48 & 6.71 & 5.57 & 6.9 & 5.38 & 0.53 & 0.42 & 0.57 & 0.4 & 7.07 & 5.06 & 7.19 & 4.86\\
	Italy & 0.71 & 0.86 & 0.79 & 0.78 & 5.25 & 6.56 & 6.03 & 5.88 & 0.68 & 0.78 & 0.73 & 0.73 & 4.92 & 5.73 & 5.33 & 5.35\\
	Lithuania & 0.89 & 0.85 & 0.93 & 0.82 & 5.34 & 5.17 & 5.59 & 4.99 & 0.95 & 0.88 & 0.96 & 0.87 & 5.61 & 5.19 & 5.6 & 5.21\\
	\addlinespace
	Luxembourg & 0.66 & 0.68 & 0.7 & 0.64 & 7.47 & 7.37 & 7.81 & 7.03 & 0.79 & 0.58 & 0.8 & 0.57 & 9.61 & 6.43 & 9.67 & 6.3\\
	Netherlands & 0.49 & 0.5 & 0.51 & 0.48 & 4.25 & 4.48 & 4.59 & 4.2 & 0.5 & 0.49 & 0.52 & 0.46 & 4.39 & 4.22 & 4.63 & 3.9\\
	Norway & 0.57 & 0.53 & 0.61 & 0.5 & 5.53 & 4.9 & 5.77 & 4.68 & 0.61 & 0.47 & 0.62 & 0.46 & 6.38 & 4.51 & 6.42 & 4.49\\
	Poland & 0.55 & 0.62 & 0.6 & 0.57 & 3.85 & 4.52 & 4.27 & 4.03 & 0.56 & 0.59 & 0.6 & 0.54 & 3.86 & 4.07 & 4.21 & 3.71\\
	Portugal & 0.86 & 0.96 & 0.94 & 0.89 & 5.96 & 6.91 & 6.52 & 6.31 & 0.87 & 0.87 & 0.9 & 0.84 & 6.03 & 6.01 & 6.11 & 5.82\\
	\addlinespace
	Romania & 0.77 & 0.87 & 0.8 & 0.85 & 4.65 & 5.44 & 4.81 & 5.36 & 0.74 & 0.79 & 0.77 & 0.76 & 4.36 & 4.9 & 4.6 & 4.68\\
	Slovenia & 0.6 & 0.66 & 0.64 & 0.62 & 4.69 & 5.18 & 4.99 & 4.93 & 0.61 & 0.61 & 0.64 & 0.58 & 4.64 & 4.68 & 4.93 & 4.36\\
	Spain & 0.66 & 0.74 & 0.72 & 0.69 & 5.45 & 6.38 & 6.12 & 5.67 & 0.64 & 0.67 & 0.68 & 0.62 & 5.3 & 5.38 & 5.63 & 5\\
	Sweden & 0.54 & 0.54 & 0.58 & 0.51 & 4.24 & 4.23 & 4.6 & 3.87 & 0.61 & 0.49 & 0.64 & 0.46 & 5.52 & 3.93 & 5.68 & 3.52\\
	Switzerland & 0.45 & 0.5 & 0.5 & 0.46 & 4.32 & 4.68 & 4.82 & 4.16 & 0.45 & 0.42 & 0.47 & 0.4 & 4.41 & 3.79 & 4.63 & 3.57\\
	\addlinespace
	No. * & 15 & 0 & 0 & 10 & 12 & 0 & 0 & 13 & 6 & 0 & 0 & 19 & 3 & 1 & 0 & 21\\
	\bottomrule
	\multicolumn{17}{l}{\rule{0pt}{1em}\textsuperscript{*} Number of countries in which the model performs the best}\\
\end{longtable}
\endgroup{}
\end{table}

\newpage\textcolor{white}{space}\newpage
\section{Accuracy of the forecasts}
\label{appendixC}
Table \ref{tab:mape2} shows how well each model could have predicted the CDRs before the COVID-19 pandemic. When using 5 years as a fitting period, the preferred model is the SP for most of the countries. However, the forecasting accuracy improves when 10 years are used to fit the models, and the SP-STFS model becomes the preferred one. According to the mean MAPE (and mean RMSE), the SP-STFS would have been more accurate in predicting mortality one year ahead in 11 cases out of 25 (and 11 cases for the RMSE) than the SP model (8 and 10 cases) and the SP-STSS model (6 and 4 cases). Therefore, for those countries, the trend is better approximated by more flexible models than a linear interpolation.
\begin{table}[h!]
\begingroup\fontsize{8}{10}\selectfont
\begin{longtable}[c]{>{}l|cc>{}c|cc>{}c|cc>{}c|clc}
	\caption{\label{tab:mape2}\textbf{Mean RMSE and MAPE on death counts in 25 European countries for multiple fitting periods based on a rolling-window scheme}. Three models (SP, SP-STSS, and SP-STFS) and two lengths of the time series are compared (5 and 10 years).}\\
	\toprule
	\multicolumn{1}{c}{ } & \multicolumn{6}{c}{5 years series} & \multicolumn{6}{c}{10 years series} \\
	\cmidrule(l{3pt}r{3pt}){2-7} \cmidrule(l{3pt}r{3pt}){8-13}
	\multicolumn{1}{c}{ } & \multicolumn{3}{c}{RMSE} & \multicolumn{3}{c}{MAPE} & \multicolumn{3}{c}{RMSE} & \multicolumn{3}{c}{MAPE} \\
	\cmidrule(l{3pt}r{3pt}){2-4} \cmidrule(l{3pt}r{3pt}){5-7} \cmidrule(l{3pt}r{3pt}){8-10} \cmidrule(l{3pt}r{3pt}){11-13}
	Country & SP & STSS & STFS & SP & STSS & STFS & SP & STSS & STFS & SP & STSS & STFS\\
	\midrule
	Austria & 368 & 375 & 370 & 4.06 & 4.12 & 4.06 & 363 & 358 & 356 & 3.98 & 3.85 & 3.84\\
	Bulgaria & 600 & 619 & 616 & 4.97 & 5.19 & 5.15 & 570 & 561 & 567 & 4.73 & 4.73 & 4.72\\
	Croatia & 285 & 295 & 289 & 4.98 & 5.2 & 5.1 & 269 & 269 & 266 & 4.35 & 4.37 & 4.39\\
	Czechia & 486 & 489 & 478 & 3.87 & 3.85 & 3.7 & 467 & 469 & 461 & 3.74 & 3.7 & 3.62\\
	Denmark & 233 & 235 & 231 & 3.88 & 3.9 & 3.86 & 227 & 225 & 219 & 3.82 & 3.65 & 3.59\\
	\addlinespace
	Estonia & 79 & 80 & 80 & 4.34 & 4.43 & 4.42 & 77 & 73 & 73 & 4.34 & 4.15 & 4.1\\
	Finland & 220 & 223 & 223 & 3.84 & 3.89 & 3.94 & 215 & 211 & 210 & 3.66 & 3.62 & 3.6\\
	France & 2,757 & 2,906 & 2,855 & 4.11 & 4.43 & 4.35 & 2,531 & 2,642 & 2,537 & 3.75 & 4 & 3.78\\
	Germany & 4,050 & 4,213 & 4,090 & 3.88 & 4.09 & 3.97 & 3,962 & 4,033 & 3,926 & 3.65 & 3.67 & 3.57\\
	Greece & 689 & 688 & 681 & 5.36 & 5.35 & 5.28 & 660 & 674 & 678 & 5.08 & 5.26 & 5.24\\
	\addlinespace
	Hungary & 662 & 691 & 680 & 4.29 & 4.58 & 4.46 & 606 & 610 & 612 & 3.8 & 3.94 & 3.94\\
	Iceland & 17 & 17 & 17 & 8.1 & 8.12 & 8.12 & 16 & 16 & 16 & 7.63 & 7.71 & 7.71\\
	Ireland & 178 & 177 & 179 & 5.33 & 5.35 & 5.47 & 164 & 156 & 156 & 4.99 & 4.76 & 4.81\\
	Italy & 3,500 & 3,805 & 3,767 & 5.33 & 5.84 & 5.79 & 3,389 & 3,633 & 3,574 & 4.99 & 5.36 & 5.27\\
	Lithuania & 216 & 212 & 207 & 5.08 & 4.94 & 4.81 & 259 & 219 & 219 & 6.2 & 5.15 & 5.16\\
	\addlinespace
	Luxembourg & 27 & 27 & 27 & 6.85 & 7.03 & 7.03 & 26 & 26 & 26 & 6.28 & 6.34 & 6.37\\
	Netherlands & 651 & 670 & 667 & 4.02 & 4.22 & 4.19 & 696 & 643 & 636 & 4.4 & 3.91 & 3.89\\
	Norway & 210 & 203 & 209 & 4.83 & 4.66 & 4.8 & 211 & 191 & 207 & 5.06 & 4.52 & 5\\
	Poland & 1,698 & 1,805 & 1,771 & 3.79 & 4.06 & 4.04 & 1,780 & 1,724 & 1,705 & 4 & 3.71 & 3.72\\
	Portugal & 731 & 773 & 771 & 5.81 & 6.32 & 6.36 & 708 & 729 & 721 & 5.58 & 5.86 & 5.8\\
	\addlinespace
	Romania & 1,396 & 1,457 & 1,458 & 5.06 & 5.3 & 5.3 & 1,244 & 1,287 & 1,304 & 4.27 & 4.68 & 4.68\\
	Slovenia & 102 & 104 & 104 & 4.88 & 4.91 & 4.92 & 99 & 99 & 99 & 4.36 & 4.36 & 4.32\\
	Spain & 2,447 & 2,581 & 2,525 & 5.42 & 5.7 & 5.52 & 2,339 & 2,416 & 2,358 & 4.93 & 5.01 & 4.9\\
	Sweden & 398 & 395 & 391 & 3.85 & 3.86 & 3.8 & 364 & 368 & 365 & 3.46 & 3.53 & 3.52\\
	Switzerland & 293 & 302 & 297 & 4.06 & 4.24 & 4.16 & 273 & 272 & 269 & 3.73 & 3.63 & 3.57\\
	\addlinespace
	No. countries * & 17 & 3 & 5 & 19 & 0 & 6 & 8 & 6 & 11 & 10 & 4 & 11\\
	\bottomrule
	\multicolumn{13}{l}{\rule{0pt}{1em}\textsuperscript{*} Number of countries in which the model performs the best}\\
\end{longtable}
\endgroup{}
\end{table}
\newpage
\section{Excess mortality during the COVID-19 pandemic}
\label{appendixD}
Tables \ref{tab:excess_counts_EU} and \ref{tab:excess_rates_EU} show the estimates of the excess deaths and rates from March 2020 to June 2022 for 25 European countries. The baseline is estimated on a common period (2010 to January 2020) using the SP-STFS model (except for Ireland, Lithuania, Norway, and Poland, for which the SP-STSS model is used). 

\begin{table}[h!]
\begingroup\fontsize{10}{12}\selectfont
\begin{longtable}[c]{llll}
	\caption{\label{tab:excess_counts_EU}\textbf{Excess death in 25 European countries}. The excess is divided into three pandemic periods: the first COVID-19 wave, and the first and second pandemic year after the shock.}\\
	\toprule
	\multicolumn{1}{c}{ } & \multicolumn{3}{c}{Excess death (prediction  intervals)} \\
	\cmidrule(l{3pt}r{3pt}){2-4}
	Country & March to June 2020 & July 2020 to June 2021 & July 2021 to June 2022\\
	\midrule
	Austria & -172 (-531 ; 183) & 8,873 (7,574 ; 10,152) & 8,154 (6,016 ; 10,238)\\
	Bulgaria & -1,954 (-2,237 ; -1,673) & 34,015 (33,203 ; 34,822) & 34,863 (33,787 ; 35,928)\\
	Croatia & -320 (-647 ; 2) & 11,647 (10,333 ; 12,927) & 13,168 (10,858 ; 15,372)\\
	Czechia & -862 (-1,418 ; -313) & 35,580 (33,060 ; 38,042) & 9,453 (4,622 ; 14,082)\\
	Denmark & -521 (-864 ; -182) & 965 (-442 ; 2,335) & 5,655 (3,126 ; 8,069)\\
	\addlinespace
	Estonia & -69 (-177 ; 37) & 1,934 (1,627 ; 2,236) & 2,504 (2,096 ; 2,901)\\
	Finland & 158 (-30 ; 344) & 325 (-197 ; 842) & 6,077 (5,413 ; 6,734)\\
	France & 18,054 (16,906 ; 19,196) & 58,582 (53,817 ; 63,310) & 42,477 (33,881 ; 50,952)\\
	Germany & -2,855 (-4,399 ; -1,318) & 69,732 (63,055 ; 76,360) & 84,476 (72,071 ; 96,716)\\
	Greece & -2,898 (-3,511 ; -2,292) & 4,018 (1,153 ; 6,819) & 12,820 (7,130 ; 18,276)\\
	\addlinespace
	Hungary & -629 (-1,178 ; -85) & 33,175 (30,842 ; 35,462) & 23,410 (19,193 ; 27,483)\\
	Iceland & -37 (-71 ; -4) & -77 (-164 ; 8) & 143 (44 ; 239)\\
	Ireland & 1,013 (828 ; 1,196) & 944 (350 ; 1,527) & 2,362 (1,479 ; 3,222)\\
	Italy & 43,900 (43,046 ; 44,752) & 105,750 (102,981 ; 108,506) & 72,607 (68,461 ; 76,726)\\
	Lithuania & 588 (374 ; 800) & 9,971 (9,271 ; 10,658) & 10,530 (9,482 ; 11,548)\\
	\addlinespace
	Luxembourg & -54 (-101 ; -8) & 284 (163 ; 401) & 34 (-104 ; 169)\\
	Netherlands & 6,340 (5,983 ; 6,696) & 10,444 (9,391 ; 11,490) & 13,677 (12,234 ; 15,106)\\
	Norway & 281 (-77 ; 632) & 410 (-769 ; 1,555) & 3,633 (1,717 ; 5,461)\\
	Poland & -3,688 (-4,426 ; -2,954) & 121,220 (118,979 ; 123,449) & 60,763 (57,545 ; 63,956)\\
	Portugal & 405 (109 ; 699) & 15,623 (14,773 ; 16,468) & 6,241 (5,102 ; 7,369)\\
	\addlinespace
	Romania & -1,545 (-2,161 ; -932) & 66,455 (64,286 ; 68,604) & 65,471 (61,995 ; 68,901)\\
	Slovenia & -80 (-265 ; 101) & 4,259 (3,578 ; 4,918) & 2,179 (1,036 ; 3,261)\\
	Spain & 42,043 (40,909 ; 43,171) & 40,576 (35,226 ; 45,855) & 35,654 (25,204 ; 45,842)\\
	Sweden & 5,582 (5,114 ; 6,045) & 5,421 (3,374 ; 7,418) & 5,097 (1,296 ; 8,735)\\
	Switzerland & 579 (246 ; 907) & 7,589 (6,344 ; 8,812) & 4,386 (2,288 ; 6,421)\\
	\bottomrule
\end{longtable}
\endgroup{}
\end{table}

\begin{table}[h!t]
\begingroup\fontsize{10}{12}\selectfont
\begin{longtable}[c]{llll}
	\caption{\label{tab:excess_rates_EU}\textbf{Excess CDRs (x 1000) in 25 European countries}. The excess is divided into three pandemic periods: the first COVID-19 wave, and the first and second pandemic year after the shock.}\\
	\toprule
	\multicolumn{1}{c}{ } & \multicolumn{3}{c}{Excess death (prediction  intervals)} \\
	\cmidrule(l{3pt}r{3pt}){2-4}
	Country & March to June 2020 & July 2020 to June 2021 & July 2021 to June 2022\\
	\midrule
	Austria & -0.19 (-0.8 ; 0.41) & 11.8 (9.17 ; 14.37) & 7.34 (5.28 ; 9.33)\\
	Bulgaria & -3.76 (-4.29 ; -3.23) & 61.45 (59.88 ; 63.01) & 45.59 (44.64 ; 46.53)\\
	Croatia & -0.99 (-2.03 ; 0.04) & 35.84 (31.49 ; 40.07) & 25.67 (22.35 ; 28.85)\\
	Czechia & -0.56 (-1.21 ; 0.08) & 42.04 (38.98 ; 45.02) & 9.96 (7.47 ; 12.36)\\
	Denmark & -0.96 (-1.7 ; -0.24) & 2.2 (-0.9 ; 5.2) & 6.97 (4.59 ; 9.25)\\
	\addlinespace
	Estonia & -0.97 (-1.8 ; -0.15) & 16.61 (14.42 ; 18.75) & 14.12 (12.93 ; 15.29)\\
	Finland & 0.52 (0.11 ; 0.94) & 1.15 (-0.02 ; 2.31) & 6.74 (6.05 ; 7.42)\\
	France & 2.92 (2.76 ; 3.07) & 9.36 (8.84 ; 9.87) & 3.41 (3.06 ; 3.76)\\
	Germany & -0.34 (-0.58 ; -0.11) & 10.2 (9.11 ; 11.28) & 8.45 (7.59 ; 9.31)\\
	Greece & -3.36 (-4.07 ; -2.65) & 4.97 (1.49 ; 8.37) & 11.86 (8.96 ; 14.66)\\
	\addlinespace
	Hungary & -0.62 (-1.33 ; 0.08) & 42.17 (39.07 ; 45.21) & 20.19 (17.78 ; 22.53)\\
	Iceland & -1.22 (-2.58 ; 0.08) & -2.25 (-6.12 ; 1.44) & -0.35 (-2.67 ; 1.83)\\
	Ireland & 2.39 (1.88 ; 2.89) & 2.14 (0.36 ; 3.87) & 3.33 (2.1 ; 4.51)\\
	Italy & 9.13 (8.96 ; 9.31) & 22.28 (21.69 ; 22.87) & 7.39 (7.01 ; 7.78)\\
	Lithuania & 2.11 (1.13 ; 3.07) & 41.56 (38.26 ; 44.8) & 28.14 (25.91 ; 30.3)\\
	\addlinespace
	Luxembourg & -1.03 (-2.09 ; 0) & 5.46 (2.5 ; 8.31) & 1.94 (0.22 ; 3.59)\\
	Netherlands & 4.83 (4.44 ; 5.22) & 8.59 (7.03 ; 10.13) & 8.39 (7.22 ; 9.54)\\
	Norway & 0.66 (-0.19 ; 1.48) & 1.01 (-1.88 ; 3.81) & 1.55 (-0.47 ; 3.48)\\
	Poland & -0.58 (-0.82 ; -0.34) & 41.48 (40.73 ; 42.22) & 16.25 (15.77 ; 16.72)\\
	Portugal & 0.08 (-0.27 ; 0.43) & 16.96 (15.93 ; 17.99) & 2.97 (2.35 ; 3.58)\\
	\addlinespace
	Romania & -0.86 (-1.26 ; -0.47) & 42.05 (40.61 ; 43.47) & 33.38 (32.37 ; 34.37)\\
	Slovenia & -0.59 (-1.66 ; 0.45) & 24.23 (20.27 ; 28.06) & 8.75 (5.91 ; 11.45)\\
	Spain & 10.59 (10.29 ; 10.88) & 10.26 (8.8 ; 11.7) & 5.03 (3.84 ; 6.19)\\
	Sweden & 6.55 (5.98 ; 7.1) & 6.45 (3.92 ; 8.93) & 3.1 (1.08 ; 5.04)\\
	Switzerland & 0.86 (0.34 ; 1.37) & 10.62 (8.53 ; 12.66) & 4.14 (2.55 ; 5.67)\\
	\bottomrule
\end{longtable}
\endgroup{}
\end{table}

\begin{table}[h!t]
\begingroup\fontsize{10}{12}\selectfont
\begin{longtable}[c]{ccccc}
	\caption{\label{tab:excess_count_AS}\textbf{Excess death in Denmark, Italy, Spain, Sweden, by sex and age}. The excess is divided into two pandemic periods: the first COVID-19 wave and the first epidemic year after the shock.}\\
	\toprule
	\multicolumn{1}{c}{ } & \multicolumn{4}{c}{Excess death (prediction  intervals)} \\
	\cmidrule(l{3pt}r{3pt}){2-5}
	Age & Denmark & Italy & Spain & Sweden\\
	\midrule
	\addlinespace[0.3em]
	\multicolumn{5}{l}{\textbf{Men}}\\
	\midrule
	\addlinespace[0.3em]
	\multicolumn{5}{l}{\textbf{March to June 2020}}\\
	\hspace{1em}\hspace{1em}0-64 & -118 (-197 ; -42) & 1924 (1755 ; 2090) & 1177 (969 ; 1383) & 284 (230 ; 337)\\
	\hspace{1em}\hspace{1em}65-74 & -122 (-218 ; -28) & 5145 (4847 ; 5439) & 3159 (2856 ; 3457) & 449 (337 ; 558)\\
	\hspace{1em}\hspace{1em}75-84 & -52 (-171 ; 64) & 9274 (8922 ; 9624) & 7253 (6854 ; 7646) & 1085 (926 ; 1240)\\
	\hspace{1em}\hspace{1em}85 & -98 (-164 ; -34) & 6900 (6481 ; 7315) & 8439 (7956 ; 8914) & 1138 (1022 ; 1253)\\
	\addlinespace[0.3em]
	\multicolumn{5}{l}{\textbf{July 2020 to June 2021}}\\
	\hspace{1em}\hspace{1em}0-64 & -143 (-409 ; 109) & 6639 (6162 ; 7110) & 2803 (2117 ; 3477) & 280 (141 ; 416)\\
	\hspace{1em}\hspace{1em}65-74 & -306 (-644 ; 15) & 14100 (12971 ; 15203) & 4331 (3037 ; 5582) & 578 (180 ; 957)\\
	\hspace{1em}\hspace{1em}75-84 & 268 (-172 ; 686) & 20323 (19141 ; 21492) & 8705 (6938 ; 10416) & 1124 (506 ; 1716)\\
	\hspace{1em}\hspace{1em}85 & 110 (-60 ; 276) & 15445 (13934 ; 16935) & 7192 (4943 ; 9374) & 739 (385 ; 1084)\\
	\addlinespace[0.3em]
	\midrule
	\multicolumn{5}{l}{\textbf{Women}}\\
	\midrule
	\addlinespace[0.3em]
	\multicolumn{5}{l}{\textbf{March to June 2020}}\\
	\hspace{1em}\hspace{1em}0-64 & -14 (-50 ; 21) & 428 (288 ; 566) & 817 (679 ; 954) & 26 (-17 ; 67)\\
	\hspace{1em}\hspace{1em}65-74 & -129 (-176 ; -82) & 1934 (1695 ; 2168) & 1318 (1198 ; 1437) & 51 (-40 ; 138)\\
	\hspace{1em}\hspace{1em}75-84 & 4 (-103 ; 108) & 5431 (5148 ; 5713) & 5664 (5333 ; 5989) & 806 (684 ; 924)\\
	\hspace{1em}\hspace{1em}85 & -303 (-385 ; -222) & 14599 (13726 ; 15463) & 13565 (12899 ; 14223) & 1506 (1292 ; 1715)\\
	\addlinespace[0.3em]
	\multicolumn{5}{l}{\textbf{July 2020 to June 2021}}\\
	\hspace{1em}\hspace{1em}0-64 & 62 (-28 ; 149) & 2454 (2040 ; 2862) & 802 (362 ; 1234) & -57 (-165 ; 48)\\
	\hspace{1em}\hspace{1em}65-74 & -191 (-313 ; -71) & 6785 (5863 ; 7682) & 1712 (1367 ; 2052) & 86 (-225 ; 382)\\
	\hspace{1em}\hspace{1em}75-84 & 94 (-289 ; 459) & 13862 (12991 ; 14723) & 7390 (5996 ; 8739) & 1056 (646 ; 1452)\\
	\hspace{1em}\hspace{1em}85 & -80 (-294 ; 130) & 28963 (24425 ; 33385) & 6234 (2922 ; 9455) & 966 (124 ; 1777)\\
	\bottomrule
\end{longtable}
\endgroup{}
\end{table}

\end{document}